\documentclass[aps, twocolumn,floatfix,pre,showpacs, superscriptaddress]{revtex4-2}

\usepackage{graphicx}
\usepackage{amsmath,units}
\usepackage{mathrsfs}
\usepackage{color}
\usepackage{xcolor}
\usepackage{appendix}
\usepackage[export]{adjustbox}
\usepackage[colorlinks=true, citecolor=cyan, urlcolor=purple,linkcolor=purple]{hyperref}
\usepackage{ulem}

\begin{document}

\title{Piston-Like Information Engine I: Universal Features in Equilibrium}

\author{R\'emi Goerlich}
\thanks{These two authors contributed equally}
\affiliation{School of Chemistry, Raymond and Beverly Sackler Faculty of Exact Sciences, Tel Aviv University, Tel Aviv 6997801, Israel}
\author{Gilad Pollack}
\thanks{These two authors contributed equally}
\affiliation{School of Physics \& Astronomy, Raymond and Beverly Sackler Faculty of Exact Sciences, Tel Aviv University, Tel Aviv 6997801, Israel}
\author{Eli Flaxer}
\affiliation{Afeka, Tel Aviv Academic College of Engineering, Tel Aviv 6997801, Israel}
 \author{Saar Rahav}
\email{rahavs@ch.technion.ac.il}
\affiliation{Schulich Faculty of Chemistry, Technion-Israel, Institute of Technology, Haifa 3200003, Israel}
\author{Yael Roichman}
\email{roichman@tauex.tau.ac.il}
\affiliation{School of Chemistry, Raymond and Beverly Sackler Faculty of Exact Sciences, Tel Aviv University, Tel Aviv 6997801, Israel}
\affiliation{School of Physics \& Astronomy, Raymond and Beverly Sackler Faculty of Exact Sciences, Tel Aviv University, Tel Aviv 6997801, Israel}
\affiliation{Center for the Physics and Chemistry of Living Systems. Tel Aviv University, 6997801, Tel Aviv, Israel}

\date{\today}

\begin{abstract}
The ability to measure the stochastic degrees of freedom of a thermal system enables the extraction of energy from an equilibrium heat bath.
This is the underlying principle of Maxwell's demon and subsequent information engines.
Here, we experimentally realize a microscopic information engine configured as a compressible piston containing a thermalized colloidal suspension.
The particle positions are recorded to identify when a predefined region near the wall is empty, allowing the piston to compress the colloidal suspension without applying work on the system.
We find that the stored compression energy is universally set by the probability of a positive measurement outcome, which in turn is controlled by parameters such as density and compression step size.
We further demonstrate that mechanical work can be extracted during the decompression of the piston, thereby closing the engine's operating cycle.
\end{abstract}

\maketitle

\textit{Introduction} \---\
Information engines employ measurements to extract additional work out of a thermal process
\cite{parrondo_thermodynamics_2015, Goerlich2025}.
Their operating mechanism, which uses measurement outcome to update our knowledge of the system's state, is fundamentally different than that of conventional heat engines.
It allows microscopic systems to convert the heat of a single heat bath into work \cite{roldan_universal_2014, koski_experimental_2014, ribezzi-crivellari_large_2019}.
This apparent thermodynamics paradox is resolved when accounting for the energetic cost of the associated information processing \cite{landauer1961irreversibility, lutz2015information}.
The framework of information thermodynamics allows a consistent description of these novel engines and shows that they do not break the Second Law \cite{esposito2011second, sagawa_generalized_2010, horowitz_nonequilibrium_2010}.
Information engines offer new possibilities to control fluctuations as well as energy and information flows through sub-parts of a system \cite{horowitz2014thermodynamics, khadka2018active, hartich2014stochastic}.
For example, it is likely that biological processes at the microscopic scale use information engine-like mechanism.
Recent works suggested that kinesin utilizes the nonequilibrium noise in cells to accelerate cargo transport \cite{ariga2021noise}, which has been interpreted as an autonomous Maxwell demon, due to reversed heat flow in the motor \cite{buisson2025hunting}.

Experimentally, information engines have been realized at the microscopic \cite{toyabe_experimental_2010, roldan_universal_2014, ribezzi-crivellari_large_2019, koski_experimental_2014, archambault2024inertial, archambault2025} and macroscopic \cite{lagoin2022human, chor2023many} scales, demonstrating the feasibility of what once was only thought experiments.
These works strengthen the link between information and thermodynamics through the use of explicit measurement and feedback loops.
Yet, on the microscopic scale of thermal energies, these experiments are still limited to single particle devices, a simplification of the complex environments of microbiological processes \cite{howard2002mechanics}.
More importantly, a critical aspect of information engine operation, the extraction of energy in the form of useful mechanical work, remains insufficiently explored due to the inherent difficulty of mechanically coupling to microscopic systems.

\begin{figure}[t]
	\centering
	\includegraphics[width=0.9\linewidth]{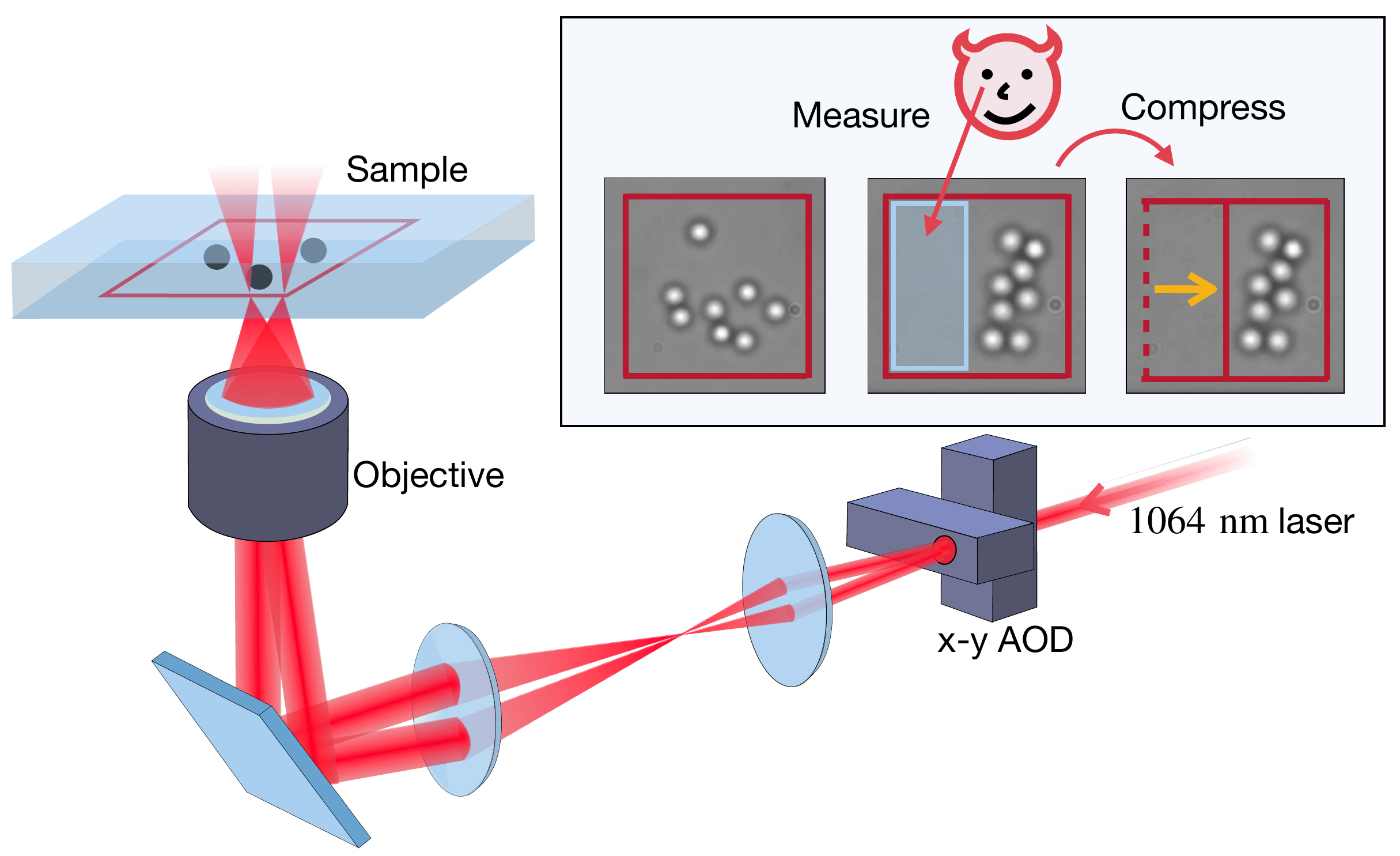}
	\caption{Schematic view of the experimental setup. A pair of accousto-optic deflectors (AODs) are used to stir a single 1064 nm laser beam at a high rate along a 2D horizontal $16 \times 16 ~\rm{\mu m^2}$ square. Colloidal particles made of silica diffusing in a high refractive index solution ($90\%~\rm{DMSO}$ and $10\%~\rm{H_20}$) are repelled by the laser beam. This setup effectively creates a bidimensional optical box enclosing the particles.
    The operation of the information engine is sketched on the right: positions of the colloidal particles are monitored in time ; if, at a time $t$, no colloidal particles are measured in an area $\Delta A$ in the vicinity of the left wall of the box, the latter can me shifted inward. Doing so, the box is compressed, increasing the osmotic pressure without applying any direct work onto the particles. To close the cycle, the energy stored in the piston can be converted back to mechanical work by allowing the colloidal suspension to expand against an obstacle, performing work on objects in its path.
    }
    \label{fig:System}
\end{figure}

Here, we realize experimentally a many-body information engine, demonstrating the extraction of energy from an equilibrium colloidal suspension and its conversion into mechanical work.
Our engine consists of several colloidal particles in a square box made of repelling optical potentials.
If one of the optical walls is moved inwards only when a predefined area near it is found to be empty, the colloidal suspension is compressed without doing any work.
This compression leads to an increase of osmotic pressure stored in the engine.
One of our main results is that the thermodynamics of this process is best described using the probability $p_1$ to find an empty area near the piston's wall. Results for different particle densities then follow a single master curve, which shows a maximum work per measurement at $ p_1^* = \frac{1}{e}$. In a companion paper, we interpret deviations from this curve as evidence for the presence of active, rather than thermal, particles in the piston.
Finally, we demonstrate the extraction of mechanical work from the compressed colloidal gas, closing the full cycle of an engine fueled by a Maxwell demon \cite{maxwell1871theory, parrondo_thermodynamics_2015, Goerlich2025}.

\begin{figure}[h!]
	\centering
	\includegraphics[width=\linewidth]{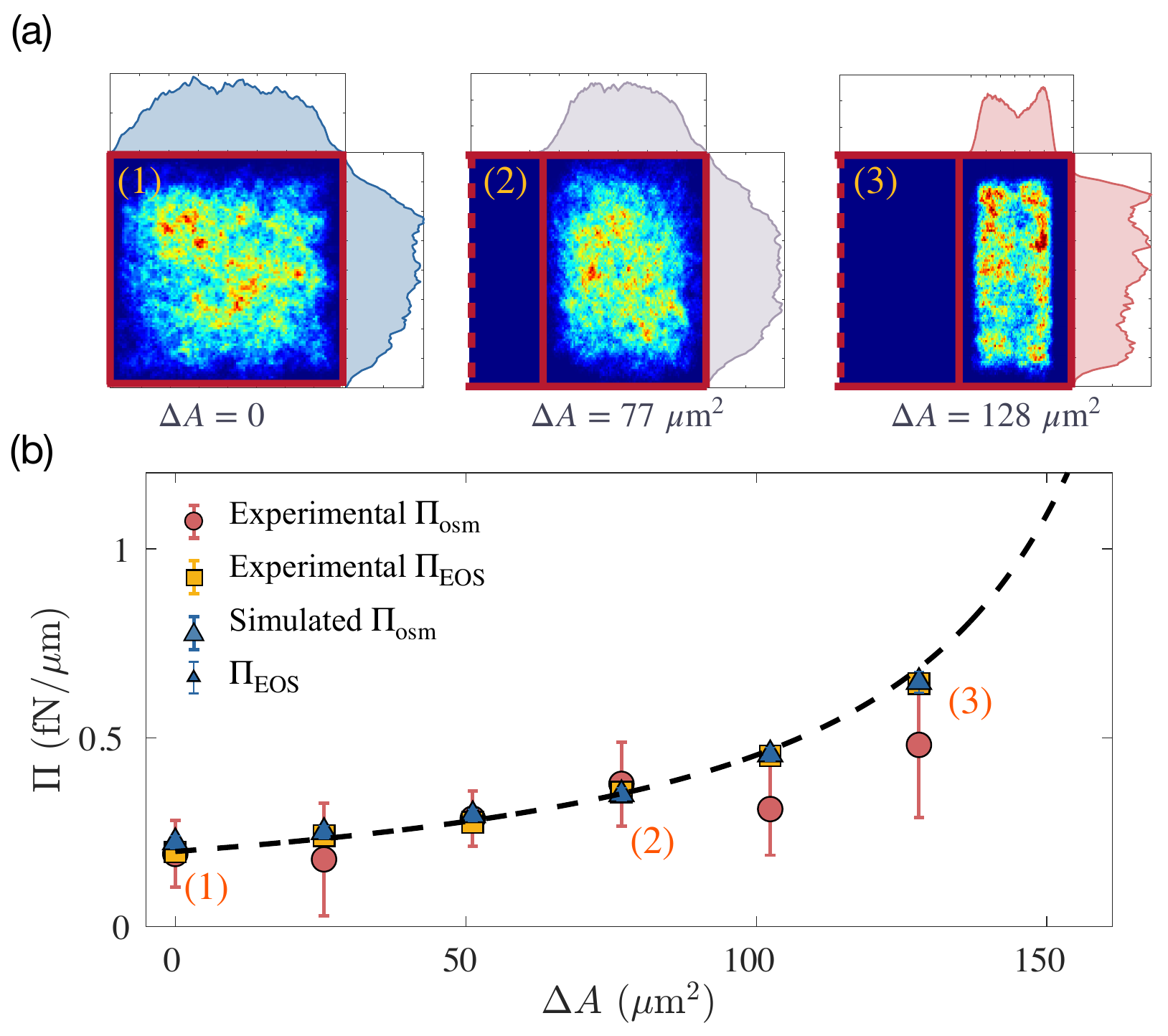}
	\caption{(a) Measured equilibrium probability densities of colloidal particles in the optical box with increasing compression $A = 16 \times 16~$ (1); $11.2 \times 16~$ (2) and $8 \times 16 ~\rm{\mu m^2} ~$(3). The shape of the optical box is superimposed in red. Along the bidimensional heatmap, we show both experimental projections along the x (horizontal) and y (vertical) axis.
    (b) Osmotic pressure evaluated in three different ways: Experimentally measured mechanical osmotic pressure $\Pi_{\rm osm}$ (red circles),  corresponding $\Pi_{\rm EOS}$ (yellow squares) calculated using the scaled particle theory approximation, mechanical osmotic pressure evaluated from numerical simulations (blue triangles), and a the exact expression of $\Pi_{\rm EOS}$ Eq.~(\ref{Eq:EOSPress}) (black dashed line). The orange numbers (1), (2) and (3) denote the box states corresponding to the densities shown in the top panel.}
    \label{fig:Pressure}
\end{figure}

\textit{Osmotic pressure increase during quasistatic compression}\---\
The working substance of our engine is an ensemble of $N=8$ colloidal particles of radius $r=1~\rm{\mu m}$ enclosed in a box of area $A_0 = 16 \times 16 ~\rm{\mu m^2}$ (the experimental setup is detailed in Appendix \ref{App:Exp}).
We study the quasistatic operation of this engine, where the system reaches thermal equilibrium between each step of compression.
To do so, we split its operation during a compression to half of its area into $6$ equidistant steps of $\Delta A = L_y \Delta x$ and study each equilibrium state independently.
After equilibration, particle trajectories are recorded during hour-long measurements.
In Fig.~\ref{fig:Pressure}~(a), we present the experimental time-averaged densities of positions $P(x,y)$ in the many-body suspension, for decreasing volumes.
The marginal densities $P(x)$ and $P(y)$ are shown alongside.
The measured densities are nearly constant near the center of the box and decay near its boundaries, revealing the softness of the optical walls.
The onset of close-packing patterns can be seen in the most compressed system (Fig~\ref{fig:Pressure} (a3)).

As the optical box is compressed, the osmotic pressure of the colloidal particles increases.
In our system, this osmotic pressure is balanced by the optical forces exerted by the confining potential. We therefore determine the pressure by measuring the mean optical force exchanged between the optical potential and the colloidal suspension. The optical force acting on each colloidal particle at a given position is evaluated as the spatial derivative of the optical potential imposed by the laser trap pattern.
The calibration of optical potential is obtained from single-particle measurements, by fitting the measured position probability density to the Boltzmann distribution, $P_1(x, y) = \mathcal{Z}_1^{-1}~e^{-U(x, y)/k_{\rm B} T}$ where $\mathcal{Z}_1$ is the partition function, and $U(x,y)$ has a Gaussian profile .
The potential $U(x,y)$, extracted from the single-particle experiment [amplitude of $5 ~\rm k_{\mathrm B}T$ and standard deviation of $0.5 ~\rm \mu m$; as detailed in Appendix \ref{App:Press}] is then used to calculate the optical force in the many-body experiments. Specifically, the force applied on particle $i^{\rm th}$ by the optical wall along its trajectory $x_i(t)$ in  the $x$-axis (and respectively the $y$-axis) is given by $f_{x,i}(t) = -\partial U[x_i(t)] / \partial x$.

The mechanical osmotic pressure corresponds to the mean (absolute value of the) forces that the optical potentials apply on the colloidal particles, divided by the length of the box walls.
The momentary force fluctuates, we therefore used
\begin{equation}
    \Pi_{\rm osm} = \frac{1}{4}\sum_{\alpha = 1}^{4} \frac{1}{L_{\alpha}} \sum_{i=1}^{8} \overline{f_{\alpha, i}(t)}
    \label{eq:MechPress}
\end{equation}
where $\overline{f_{\alpha, i}}(t)$ is the time-averaged force exerted on the $\alpha$-th wall by the $i$-th particle. 
Further details on the evaluation of $\Pi_{\rm osm}$ are provided in Appendix \ref{App:Press}.
The pressure measured in the experiment (red circles) and in numerical simulations (blue triangles) are shown in Fig.~\ref{fig:Pressure}~(b) as a function of box area.

The evaluation of $\Pi_{\rm osm}$ is complemented by the pressure derived from the scaled particle theory equation of state (EOS) for a bidimensional hard-disk fluid \cite{helfand1961theory, thorneywork2017two, royall2024colloidal}
\begin{equation}
    \Pi_{\rm EOS} = \frac{1}{(1-\phi)^2} \rho k_{\rm B} T,
    \label{Eq:EOSPress}
\end{equation}
where $\phi = N \pi r^2 / A$ is the area fraction of the colloidal particles and $\rho = N / A$ is the number density.
$\Pi_{\rm EOS}$ is shown in Fig.~\ref{fig:Pressure}~b (yellow squares and black dashed line) in good agreement with the osmotic pressure.

We note that the optical box is built of soft Gaussian walls, making the definition of $A$ subtle, impacting the values of $\phi$ and $\rho$. To overcome this ambiguity, we approximate the area of the box by the area which includes all positions in which the occupation probability is within $ 99.5\% $ of that of the center of the box (as detailed in Appendix \ref{App:Phi}).
The effective area $A$ used to derive $\Pi_{\rm EOS}$ agrees well with the osmotic pressure measurement (Fig.~\ref{fig:Pressure}~(b)).
The choice of effective area is further justified since it leads to the equality of the measured particle density in the bulk $N \langle P(x,y) \rangle$ and the theoretical number density $\rho = N/A$ (see Fig.~\ref{fig:Phi}, Appendix \ref{App:Phi}).

\begin{figure}[h!]
	\centering
	\includegraphics[width=\linewidth]{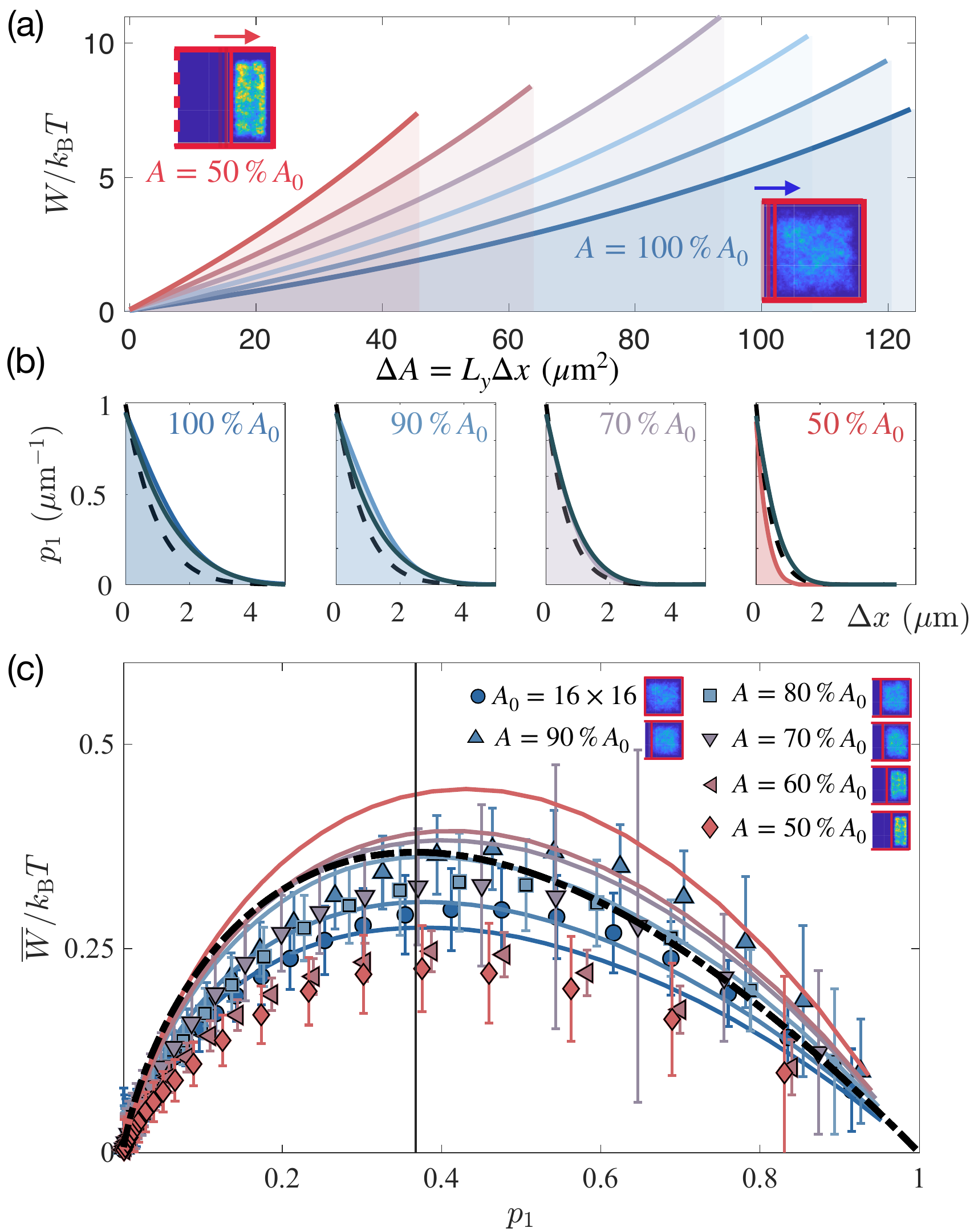}
	\caption{(a) Work $W$ stored under the form of osmotic pressure as a function of the area compression $\Delta A = L_y \times \Delta x$ for various initial box sizes (from $A_0 = 16\times16 ~\rm \mu m^2$ in blue, to $8\times16 ~\rm \mu m^2 = 50\% A_0$ in red).
    (b) Probability $p_1$ of a positive measurement outcome \textit{i.e.} that the area $\Delta A$ is empty, as a function of the width $\Delta x$ of the probes area. Experimental result (blue to red solid lines and filled area) in close agreement with numerical simulations (gray solid lines); both stay similar to the result of an ideal gas (black dashed line) $p_1^{\rm ideal} = (1 - \Delta_x/L_x)^N$.
    (c) Normalized mean work per measurement $\overline W= p_1 W$ for each box sizes, plotted against $p_1$. The experimental results (symbols) and numerical simulations (solid lines) bundle around the universal curve Eq.~(\ref{eq:universal}) (black dash-dotted line). The two most compressed cases depart more significantly.} 
    \label{fig:Work}
\end{figure}

\textit{Universal expression for the mean stored energy per measurement}\---\
The increase in the suspension's osmotic pressure can be used to extract work from the expansion of the piston.
The energy stored by the increase in pressure in colloidal suspension upon a compression $\Delta A = L_y \times \Delta x$ reads
\begin{equation}
	W(\Delta A) =\int_0^{\Delta x} \Pi(x) L_y  dx
    \label{eq:work}
\end{equation}
where we use the expression Eq.~(\ref{Eq:EOSPress}) of $\Pi_{\rm EOS}$ (see Fig.~\ref{fig:Pressure}~b) for the pressure $\Pi (x)$.
In Fig.~\ref{fig:Work}~(a), we show  $W$, the increasing osmotic energy stored by the engine, which can later be converted to work, as a function of the compression $\Delta A$.
For instance, when the full-size box $A_0$ is compressed by $50\%$ of its size, a work $W_{\rm stored} = \int_0^{L/2} \Pi(x) L_y dx = 7.26 \pm 0.46 ~\rm{k_{\mathrm B} T}$ is stored in the piston.
When compression is applied to a box of smaller initial size (blue to red lines), the slope of $W(\Delta A)$ is larger, due to the higher initial density.

In order to operate as an information engine, the compression of the piston by $\Delta x$ is conditioned by the absence of any colloidal particle in $\Delta A$ upon measurement.
This event has a finite probability $p_1$ that depends on both the size of the probed region and the density of particles.
The complementary probability $p_0 = 1-p_1$ is the probability that $\Delta A$ is populated by at least one particle.
The information processed by a single measurement reads $I = -p_0 \ln p_0 -p_1 \ln p_1$ (Appendix \ref{App:Stat}) \cite{Shannon1949, maruyama_colloquium_2009, parrondo_thermodynamics_2015, lutz2015information}.
To evaluate $p_1$, we measure the time-average probability density $P_{\rm extr}(x)$ (normalized, $\int_0^L P_{\rm extr}(x)dx = 1$) of position of the leftmost particle.
Its position delineates the largest empty area along the left wall of the box.
The probability $p_1(x)$ then reads
\begin{equation}
    p_1(\Delta x) = \int_{\Delta x}^{L}P_{\rm extr}(x') dx'
\end{equation}
and represents the probability that all degrees of freedom of the system are restricted to $x'\in[\Delta x, L]$, leaving $\Delta x$ empty.
The probability $p_1$ measured experimentally is shown in Fig.~\ref{fig:Work}~(b).
It is a decreasing function, ranging from $p_1 = 1$ for infinitesimal compression to $p_1 = 0$ for compressions of a few microns.
The decay of $p_1$ is faster as the box gets smaller (from blue to red graphs).

We now turn to the connection between the probabilistic properties of the engine, captured by $p_1$, and its thermodynamics.
On average, a number $\langle n\rangle = 1/p_1$ of measurements is performed before finding an empty $\Delta A$ and triggering one step of compression.
The mean work extracted from a single measurement, therefore, reads $\overline W= p_1 W $.
For constant rate $r$ measurements, $\overline W$ reflects the power $\mathcal{P} = r \overline W$ of the engine \textit{i.e.} the average rate of energy storage. 
Furthermore, the probability $p_1$ can be expressed as $p_1 = \mathcal{Z'}/\mathcal{Z}$ where $\mathcal{Z'}$ is the partition function of the system restricted to $x'\in[\Delta x,L]$.
In isothermal quasistatic processes, the ratio of partition functions is often interpreted as an equilibrium free-energy difference \cite{jarzynski2012equalities, gavrilov2017direct}. In our case, the partition function $\mathcal{Z}'$ in $p_1$ corresponds to inserting a virtual hard wall at $x=\Delta x$, whereas $\Delta F$ is defined from shifting the actual soft optical wall. Because the optical confinement is sufficiently steep, this mismatch is small, allowing the approximation $\Delta F \approx k_{\rm B}T \ln\left(\mathcal{Z}/\mathcal{Z'}\right) = k_B T \ln(p_1)$.
The quasistatic mean work per measurement, $\overline{W}=p_1 W$, then obeys
\begin{equation}
    \overline W \approx -k_{\mathrm B} T p_1 \ln(p_1)
    \label{eq:universal}
\end{equation}
which uniquely relates $\overline{W}$ to the success probability $p_1$.


Eq.~(\ref{eq:universal}) is an approximate universal relation between the extractable work stored in the piston per measurement, at a single compression step. It is expressed solely through the probability $p_1$.
Namely, for hard walls, the mean work per measurement reaches a maximum of $ k_{\mathrm B} T/e$ for $p_1 = 1/e$, irrespective of details such as the particle density or the strength of inter-particle interactions. This result suggests a natural approach to determining how to design the measurement (i.e., choosing $\Delta x$) used in these types of information engines.
In Fig.~\ref{fig:Work}~(c), we show $\overline W$ as a function of $p_1$ for various box sizes for experiments and simulations. All data from the soft-wall system cluster, within experimental error, around the master curve of Eq.~(\ref{eq:universal}).

\textit{Converting the osmotic pressure into mechanical work:}
Energetically, the osmotic pressure corresponds to the mean force applied by the particles against the optical boundaries of the box.
This is in line with microscopic engines using colloidal systems where the extracted energy is measured as the stochastic thermodynamic work \cite{sekimoto1998langevin} exerted against the optical potential \cite{roldan_universal_2014, ribezzi-crivellari_large_2019}.
The progressive compression corresponds to the storage of an increasing potential energy, which was previously demonstrated by gravity \cite{saha_maximizing_2021} and fluid flow \cite{admon_experimental_2018}.
To close the cycle of operation of the machine, this potential energy must be converted into useful mechanical work via an expansion process \cite{szilard1929, parrondo_thermodynamics_2015}.

\begin{figure}[h!]
	\centering
	\includegraphics[width=\linewidth]{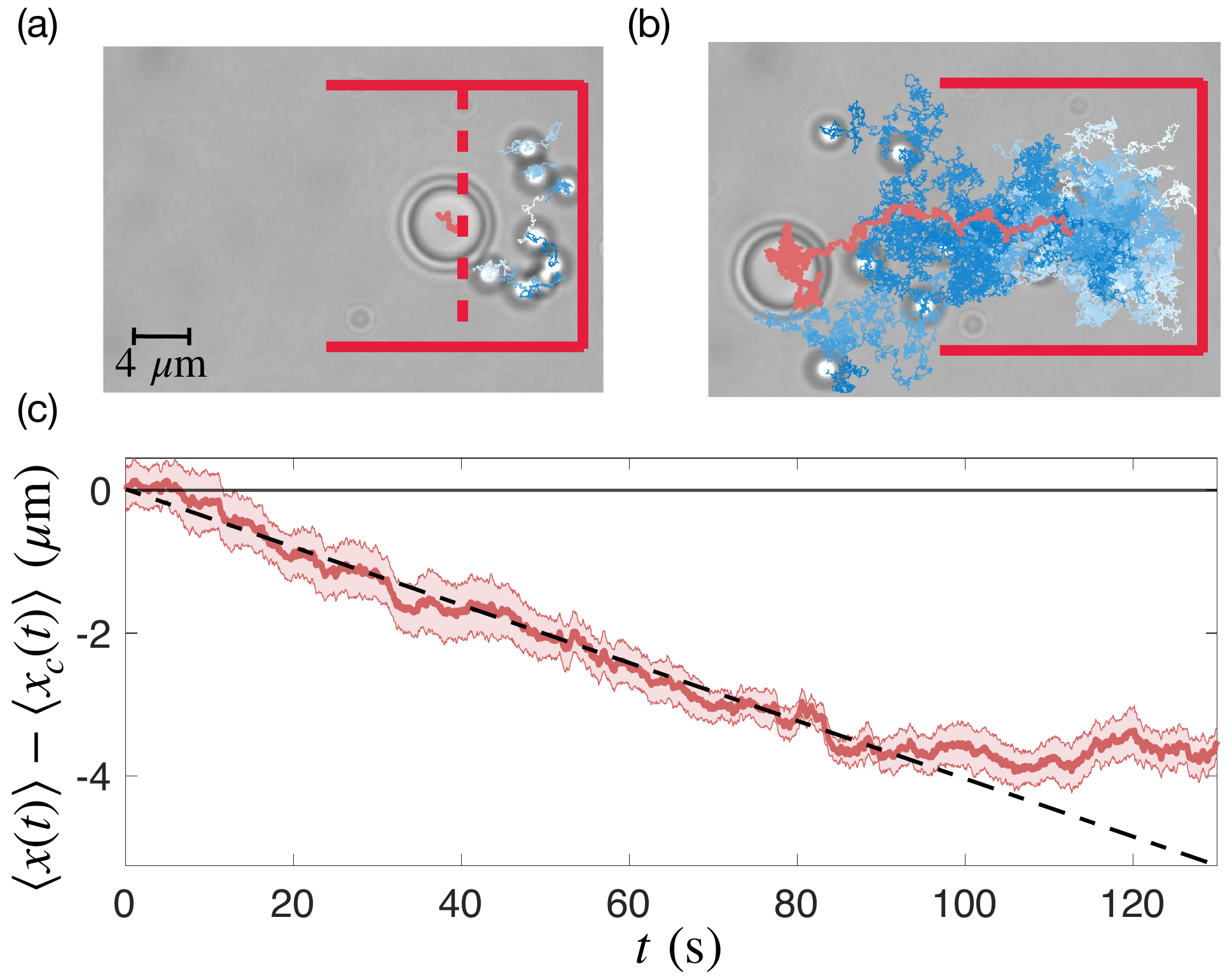}
	\caption{Converting osmotic pressure into mechanical work.
    (a) Initial state: A large diameter $d = 5 ~\rm \mu m$) colloidal particle is brought close to the wall of the compressed box, which contains $8$ colloidal particles in equilibrium.  The compressed left optical wall is then removed, letting the colloidal particle suspension to expand by freely diffusing out of the left aperture of the box. 
    (b) Final state: after a time $t \approx 8$ minutes, the small colloidal particles have escaped the confining box, pushing the large colloid along the x-axis.
    (c) Quantitative measurement of the displacement of the large colloidal particle. The drift $\langle x(t) \rangle$ is the average over 6 repetition of the experiment. The same number of control experiments are realized with the same box geometry (which applies a force on the large colloidal particle) without the enclosed small colloids. The represented drift (red line) is the difference between the experiment and the control.}
    \label{fig:Extraction}
\end{figure}

Here, we couple our micro-engine to an auxiliary mechanical system, a $5~\rm{\mu m}$-large colloidal particle, and transform the osmotic pressure into directed mechanical motion (see Fig.~\ref{fig:Extraction}~(a)).
The left optical wall is then removed, letting the enclosed colloidal particle escape the box (Fig.~\ref{fig:Extraction}~(b)).
During this free expansion of the colloidal gas, the stored pressure is released and the osmotic energy content $W_{\rm stored}$ (maximal value of the blue line shown in Fig.~\ref{fig:Work}~(a)) is made available.
The movement of the small colloidal particles displaces the larger obstacle, effectively converting $W_{\rm stored}$ into mechanical motion.
We measure the mean drift $u(t) = \langle x(t) \rangle$ of the large particle. This velocity is a result of the force applied on it by the colloidal particles which is given by the drag force $f_{o} = \gamma_{o} \dot u$ with $\gamma_{o}$ the Stokes friction coefficient of the obstacle particle.
The  associated work extracted during a time $\tau$ reads $W_{\rm out}(\tau) = \int f_{o} dx = \gamma_{o} \int_{0}^{\tau} \dot u(t)^2 dt$.
A control experiment is performed in the same configuration, but in the absence of the small colloidal particles to account for the effect of the geometric constraint imposed by the three remaining walls on the mean drift $\langle x_{\rm c}(t) \rangle$.
The extracted work is measured on the difference $\langle x(t)\rangle - \langle x_{\rm c}(t)\rangle$, ensuring that we measure only the effect of the colloidal gas expansion.

As shown in  Fig.~\ref{fig:Extraction}~(c), the particle average drift is linear over the first $\approx 100$ seconds. 
The associated extracted work is measured to be $W_{\rm out} = 3.16 \pm 0.12~\rm{k_{\mathrm B}T}$, which is close to half of the total available osmotic work $W_{\rm stored}$ (shown in Fig.~\ref{fig:Work}~(a)).
This experiment demonstrates the conversion of the energy stored by an information engine into mechanical work at the microscopic scale.
It also brings colloidal model experiments closer to biological micro-motors as the displacement of the obstacle particle is analogous to cargo transport \cite{ariga2018nonequilibrium}.\\

\textit{Conclusion:}\---\
In this work, we realize a microscopic many-body information engine consisting of an optical piston containing a colloidal suspension.
We measure the osmotic pressure and work, and show that for an information engine with thermal particles (and stiff walls), the engine's thermodynamics can be characterized solely in terms of the probability $p_1$. We use experiments and simulations to show that this simple description captures the behavior of real information engines.
Finally, we demonstrate the conversion of information to mechanical work experimentally by closing the engine's cycle.

Our results provide an experimental step toward real-world systems, exemplified by the operation of molecular motors such as kinesin \cite{ariga2018nonequilibrium}.
Like our experimental system, kinesin transport takes place in a highly fluctuating environment and operates as an information machine \cite{buisson2025hunting}.
Unlike our setup, however, kinesin functions in a strongly non-thermal environment \cite{guo2014probing}.
Such non-thermal conditions have recently been shown to enhance the efficiency of both heat engines \cite{krishnamurthy2016micrometre} and information engines \cite{chor2023many, saha2023information}.
This raises a natural question: how would the information engine studied here, which relies on number fluctuations that are typically amplified by activity \cite{ramaswamy2003active}, perform under non-equilibrium conditions? In a companion paper, we study a similar information engine with macroscopic active particles and show that the activity of the particles is expressed through qualitative differences in the shape of the $\overline{W}(p_1)$ curve.

\acknowledgements
We thank Kristian S. Olsen for insightful discussions. Y.R. and R.G. acknowledge support from the European Research Council (ERC) under the European Union’s Horizon 2020 research and innovation program (Grant Agreement No. 101002392). S.R. is grateful for support from the Israel Science Foundation (Grant No. 1929/21).\\

\appendix

\section{Experimental Optical Piston}
\label{App:Exp}
\begin{figure}[h!]
	\centering
	\includegraphics[width=\linewidth]{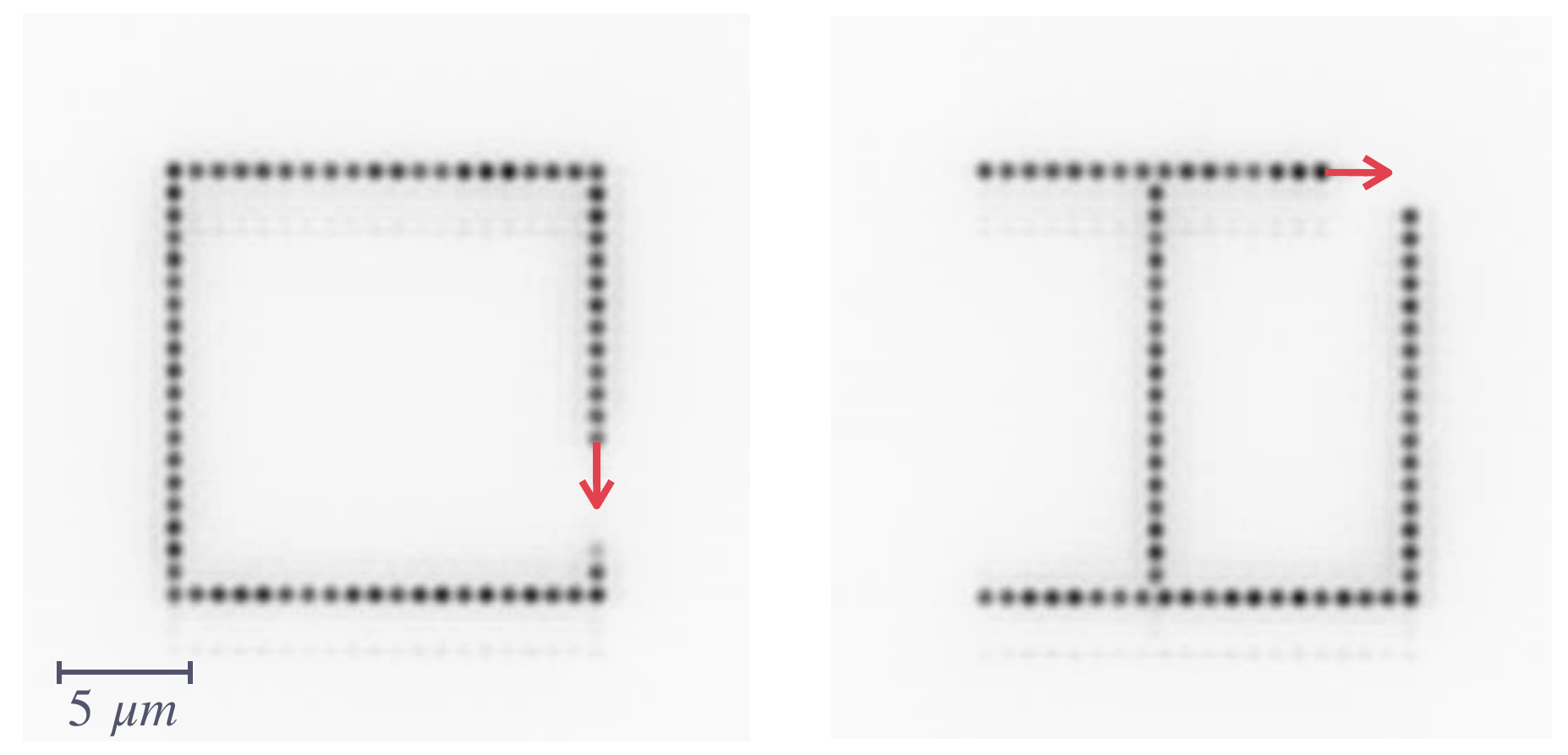}
	\caption{Optical pattern corresponding to the full (left) and partially compressed (right) optical box created by AOD beam multiplexing. A single trap in sequentially illuminating each point along a $16\times16~\rm \mu m$ square, with 20 traps per side. The red arrow is indicating the position and direction of movement of the trap illuminated last.}
    \label{fig:Pattern}
\end{figure}

The optical box is created using a single $1064$ nm laser beam (Laser Quantum OPUS 1064), which is steered by a pair of acousto-optic deflectors (AA opto-electronics 
DTSX-400-1064) controlled by a radio-frequency driver (AA opto-electronics 2-channels Direct Digital Synthesizer).
The laser beam is sent sequentially along a discrete series of locations, along the 4 walls of the box in the $x$-$y$ plane normal to the optical axis.
We use a total intensity of $300$ mW  which is shared between each trap composing the optical box (20 traps per wall, \textit{i.e.} approx $4$ mW per trap).
The resulting optical pattern is magnified using a pair of lenses, in a 4-f geometry connecting the output of the second AOD to the back-focal-plane of the microscope objective (Olympus UPlan SApo $100$x, NA 1.4, oil immersion).
A snapshot of the obtained optical pattern is shown in Fig.~\ref{fig:Pattern} both for a full square box, and for a partially compressed box.
The red arrow highlights the current location of the trap, moving along the box.
Such an optical pattern appears, over the space and time scales of the diffusing particles, as a continuous optical box.
A numerical calibration is used to further adjust the intensity of each trap location, to correct for the uneven output of the AOD.\\

The working substance is composed of silica microspheres (PolyScience Silica microspheres $d = 2.1 \pm 0.1~\rm \mu m$), diluted in a solution composed of $90\%$ of DMSO and $10\%$ water.
The index of refraction of the silica particles $n_{\rm p} \approx 1.42$ is therefore lower than the one of the solution $n_{\rm s} \approx 1.46$, ensuring a repulsive optical force.
The colloidal solution is enclosed in a fluidic cell made of a spacer of thickness $120~\mu m$ and width $1~\rm cm$ (Grace Bio-lab SecureSeal) between a glass slide and a cover-slip.
The sample is inserted horizontally above the microscope objective.
The colloidal particles sediment at the bottom of the fluidic cell and are trapped in the $x$-$y$ plane, by the repelling optical box.
The colloidal particles enclosed in the box are imaged using bright-field illumination.
We record 40 to 97 minutes long movies (longer movies are recorded for lower densities) at a rate of $15~\rm{Hz}$ with a CMOS camera (FLIR Grasshopper GS3-U3-23S6M-C).
Their trajectories are extracted with standard particle tracking algorithms \cite{crocker1996methods}.

\section{Force Calibration and Osmotic Pressure}
\label{App:Press}

The engine’s thermodynamics are governed by the osmotic pressure exerted by the colloidal suspension on the optical walls.
While the bulk osmotic pressure of hard-sphere colloids is well characterized \cite{helfand1961theory, thorneywork2017two, royall2024colloidal}, the optical confinement used here introduces both benefits and complications.
Proper calibration enables direct force measurements on each wall, yet the soft potential profile of the walls makes the definition of the effective box area ambiguous.
\begin{figure}[h!]
	\centering
	\includegraphics[width=\linewidth]{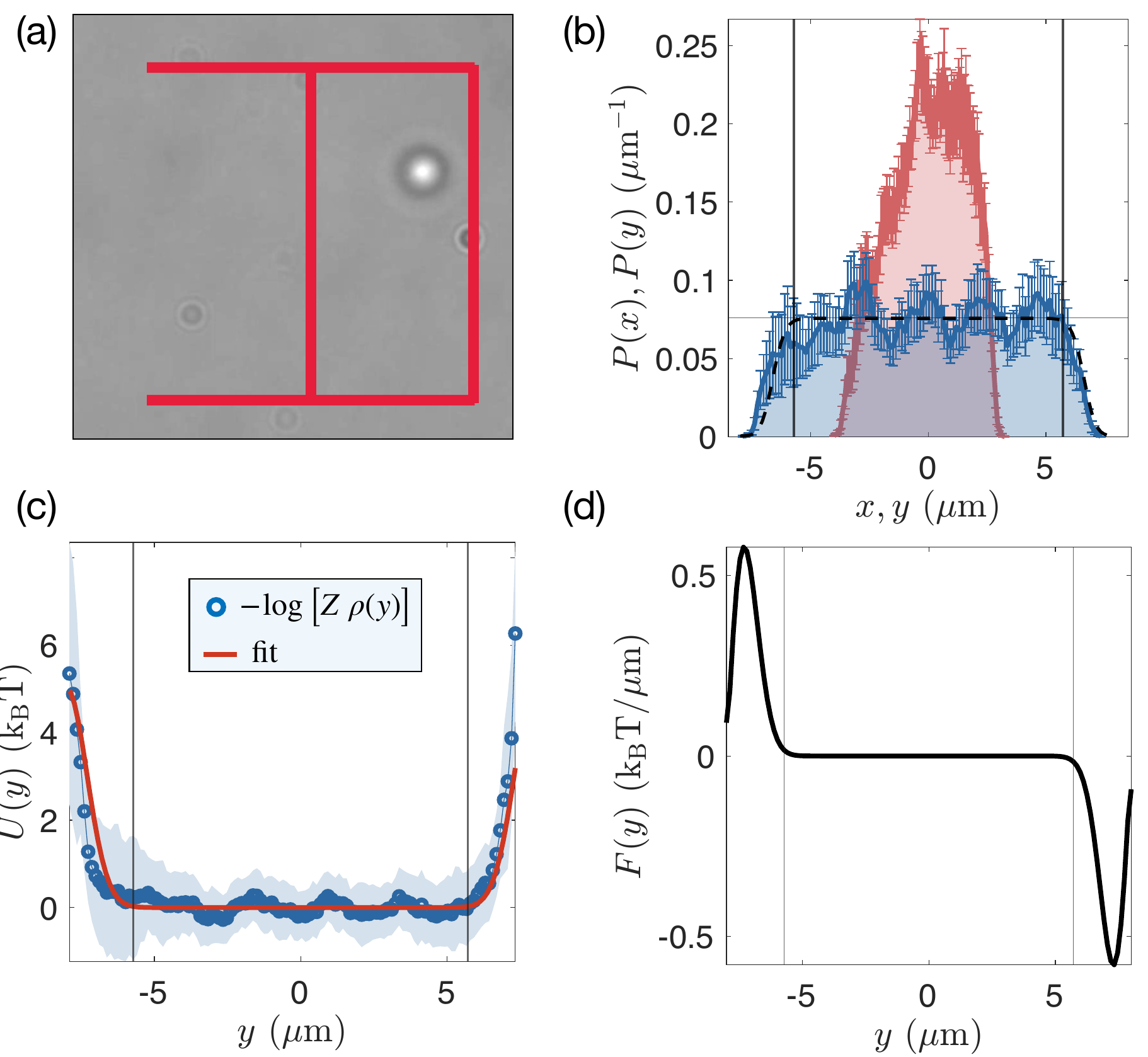}
	\caption{Measurement of the optical box potential via the diffusion of a single colloidal particle.
    (a) Snapshot of a single colloidal particle of diameter $d = 2 \rm \mu m$ diffusing in a half-compressed optical box.
    (b) Measured probability densities along the compressed x-axis (red) and the uncompressed y-axis (blue). Vertical lines underline the extension of the ``free'' inner part of the box (defined as the y-axis point where the histogram reaches the central uniform value, up to a $0.5 \%$ threshold).
    (c) Potential $U(y)$ measured (blue circles) as the log of the equilibrium Boltzmann state $P(y)$ times its partition function $\mathcal{Z}_1$. We fit this potential with the expected pair of Gaussian function $g(y) = \mathcal{A}~ e^{-(y-y_0)^2/2\sigma^2}$ with center $y_0 = \pm 8 ~\mu m$ and fitting parameters $\mathcal{A}$ and $\sigma$. The result (red solid lines) agrees with the experimental measurement within the statistical errors.
    (d) Resulting force $F(y)$ which is used throughout this work, to model the walls of the optical box.}
    \label{fig:Single}
\end{figure}

In order to characterize the optical potential, which in turn allows to determine the forces, as well as to define an effective length of the box, we study the properties of a single particle diffusing in an optical box  with $L_x = 8~\rm{\mu m}$ and $L_y = 16~\rm{\mu m}$ (Fig.~\ref{fig:Single}~(a)).
The trajectory is recorded for one hour at a frame rate of $15~\rm{s}^{-1}$.
The recorded densities along both axis are shown in Fig.~\ref{fig:Single}~(b)
We use the density along the $y$-axis to define the effective length of the box.
To do so, we set a threshold at $5\%$ of its mean central plateau value (underlined by a black horizontal line in Fig.~\ref{fig:Single}~(b)).
The $y$-value where the density crosses this threshold defines the effective length of the box (underlined by the two vertical lines).
This corresponds to considering a width of $2.15~\rm{\mu m}$ near the center of each trap as belonging to the soft optical wall.
It reduces the $16~\rm{\mu m}$-wide box to an effectively $\approx 12 ~\rm{\mu m}$-wide box, where we can consider that no optical force is applied onto the particles.
The single-particle density, as a Boltzmann equilibrium state, further allows us to obtain an estimation of the optical potential, via the logarithm of the density (Fig.~\ref{fig:Single}~(c).
It is numerically fitted with a pair of Gaussian functions, centered in $\pm 8~\rm{\mu m}$ as imposed by the laser pattern.
This is our measurement of the amplitude $\mathcal{A} = 5.03 \pm 0.2 ~\rm{k_{\rm B}T}$ and standard deviation $\sigma = 0.49 \pm 0.04 ~\rm{\mu m}$ of the Gaussian walls of the optical box.
These values are then used, throughout this work, to evaluate the force exchanged between each particle and a nearby wall.
The two vertical lines underline the same effective length of the force-free box, as detailed above.
The optical force, gradient of the Gaussian potential is shown in Fig.~\ref{fig:Single}~(c).

\begin{figure}[h!]
	\centering
	\includegraphics[width=\linewidth]{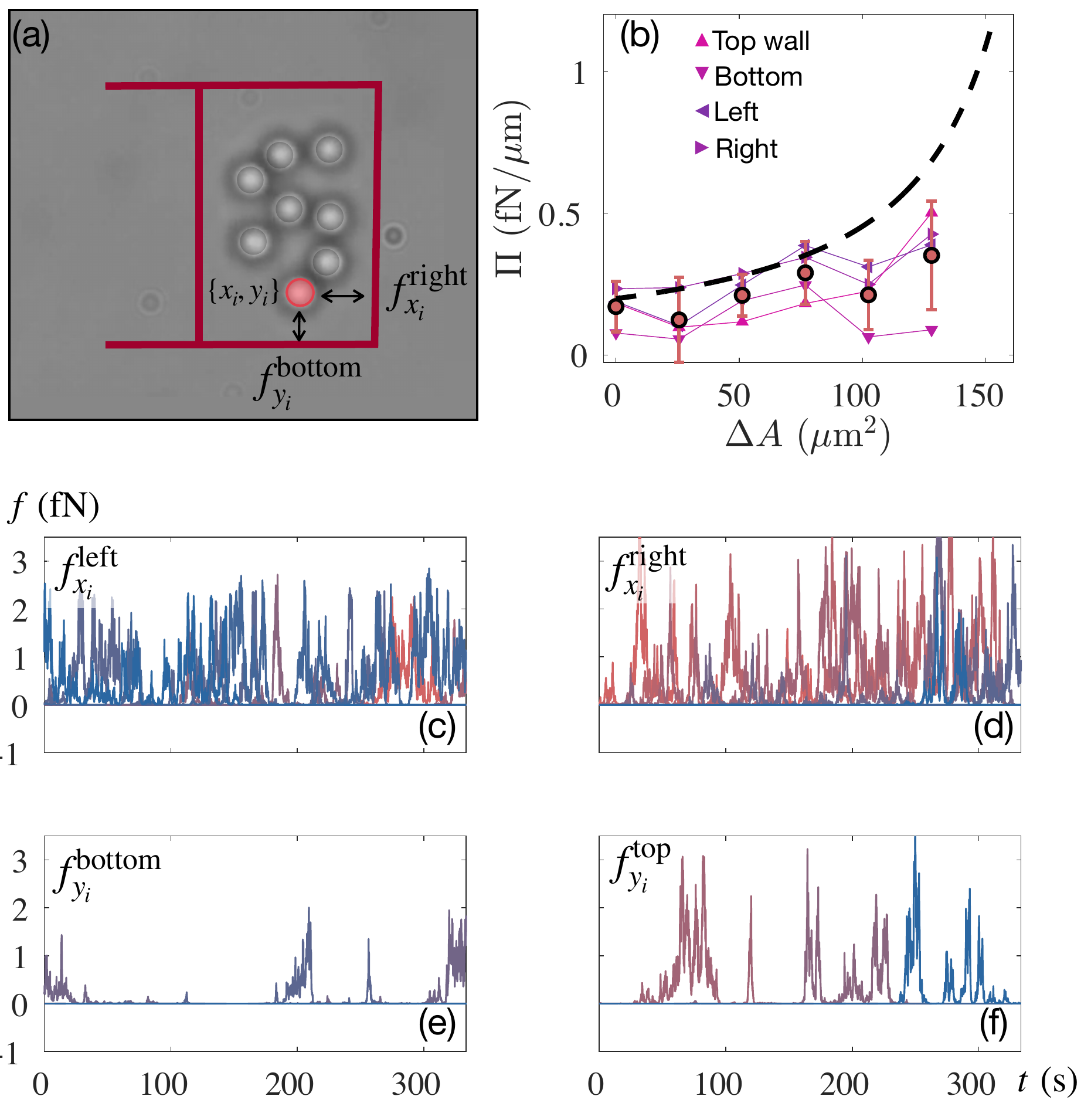}
	\caption{Details on the measurement of the mechanical osmotic pressure.
    (a) snapshot of a configuration of the box, for a given particle with $\{ x_i, y_i\}$, a force is exerted to each nearby wall.
    (b) The pressure $\Pi_{\rm osm}$  as a function of the compressed area $\Delta A$ is shown independently along each wall (pink to purple triangles) as well as the average over the four walls (red circles).
    (c,d,e,f) 300 second long portions of the time-dependent force $f(t)$ exerted against each wall, by each particle (blue to red lines), in the most compressed state ($8\times16~\rm{\mu m^2}$).
    }
    \label{fig:DetailsPresure}
\end{figure}

In Fig.~\ref{fig:DetailsPresure}, we detail the measurement of $\Pi_{\rm osm}$ as the average mechanical force exerted by all the particles against each wall, divided by the length of the respective wall.
In such an equilibrium system, we expect each wall to measure the same pressure.
As seen in Fig.~\ref{fig:DetailsPresure}~(b), the pressure measured along the different wall clusters around an increasing curve, close to the pressure evaluated via an effective EOS, Eq.~(\ref{Eq:EOSPress}) (black dashed line).
Deviations are stronger for the two most compressed states, which we can associate with the onset of closed-packing pattern formation.
Fig.~\ref{fig:DetailsPresure}~(c,d,e,f) shows that the forces experienced by each particle has a high variability: it is nearly zero when the particle diffuses in the bulk and strongly increases when the particle is approaches a wall.
The reduced length of the horizontal walls reflects in the reduced force along the $y$-axis. This however leads to approximately the same pressure when dividing by the associated wall length (Fig.~\ref{fig:DetailsPresure}~(b)).

\section{Numerical Simulations}
\label{App:Sim}
The simulations followed the standard numerical scheme for quasi-2D colloidal suspensions \cite{volpe_simulation_2013}. The inter-particle interactions were modeled using the WCA potential \cite{hess_thermomechanical_1998}, while the walls were modeled as a repulsive Gaussian potential, using the parameters measured from the experiment.
The time step used for the simulations was $\Delta t = 10^{-6}$s, and the particle positions were recorded every $0.05$s. The steady state distribution of the particles in each set of initial conditions (\textit{i.e} box size and number of particles) was obtained by averaging over 100 simulations, each running for $10^9$ steps, corresponding to 1000 seconds simulated experiment time.

\section{Effective Size and Area Fraction}
\label{App:Phi}

\begin{figure}[h!]
	\centering
	\includegraphics[width=\linewidth]{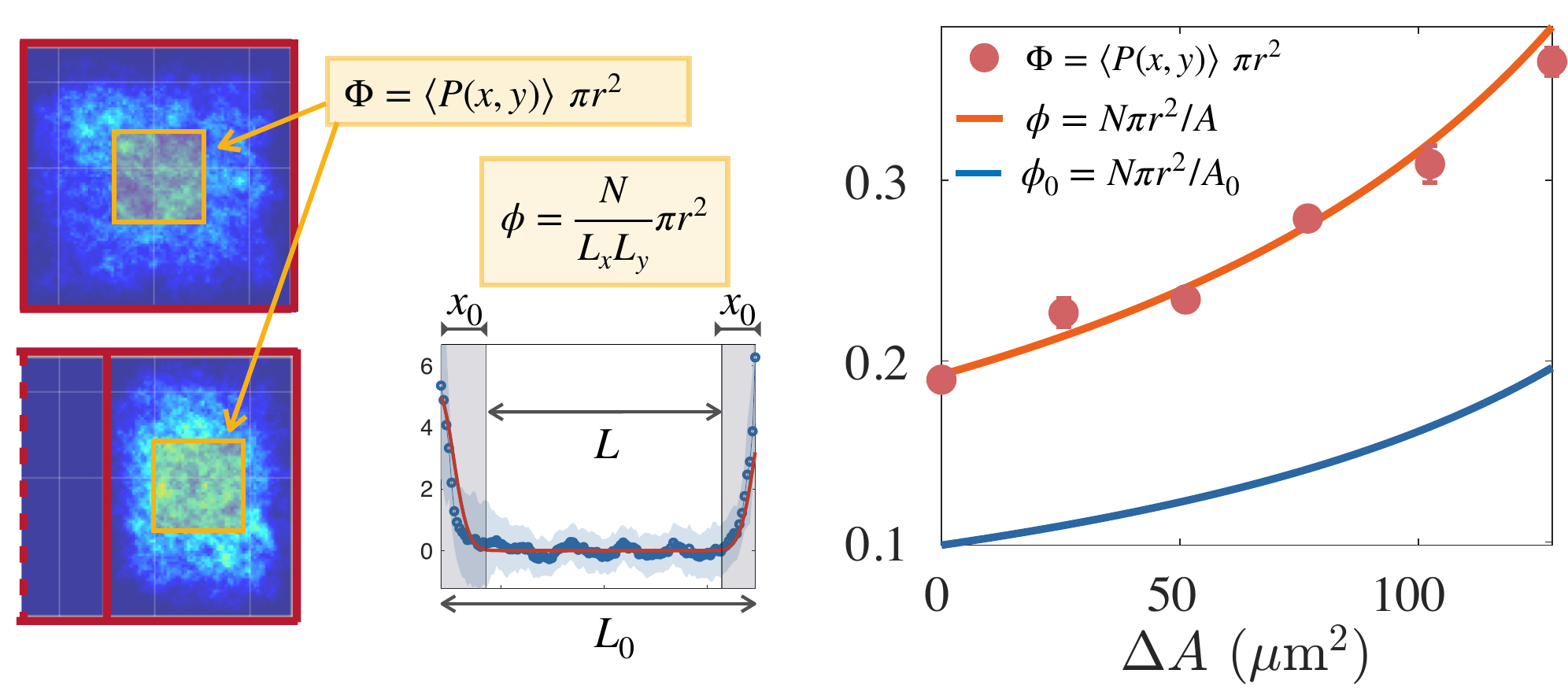}
	\caption{ (left panels) two distinct measurements of area fraction (1) via the measured mean density in a $(1.5 \rm{~to~} 2.5)\mu m$-sided area in the center of optical box ; (2) via the ratio of the occupied area $N \pi r^2$ to the effective area, defined with the threshold $x_0$.
    (right panel) area fraction $\Phi$ measured via mean density (red circles) and $\phi$ via the effective area of the optical box (red solid line); the area fraction measured on the full $\phi$ area of the box, delineated by the center of the optical potentials (blue solid line).}
    \label{fig:Phi}
\end{figure}

The calibration of the optical potential is performed using a recording of a single colloidal particle. From this recording, we also define an effective box size, $L = L_{0} - 2 x_0$ considering a width of $x_0 = 2.15~\rm{\mu m}$ near the center of each trap as belonging to the soft optical wall.

In order to corroborate the effective size of the box, we measure the average density of particles in the many body experiment for each box size, and compute the empirical area fraction $\Phi = \pi r^2 \langle P(x,y)\rangle$ with $r$ the radius of the colloidal particles.
The mean density is measured in the center of the box, on square areas of side $l\in[1.5, 2.5]~\rm{\mu m}$, and its variation for different $l$ gives a measure of the error on its estimation.
This area fraction is compared with the theoretical value $\phi_{0} = 8\pi r^2 / A_{0}$ and $\phi = 8\pi r^2 / A$.
The agreement, shown in Fig. \ref{fig:Phi} between $\Phi$ and $\phi$ shows the sensible estimation of the effective size of the box. $\Phi$ (or $\phi$) then allows to measure $\Pi_{\rm EOS}$ which agrees with $\Pi_{\rm osm}$ derived from the force applied by the particles against the optical walls.

\section{Information}
\label{App:Stat}

\begin{figure}[h!]
	\centering
	\includegraphics[width=\linewidth]{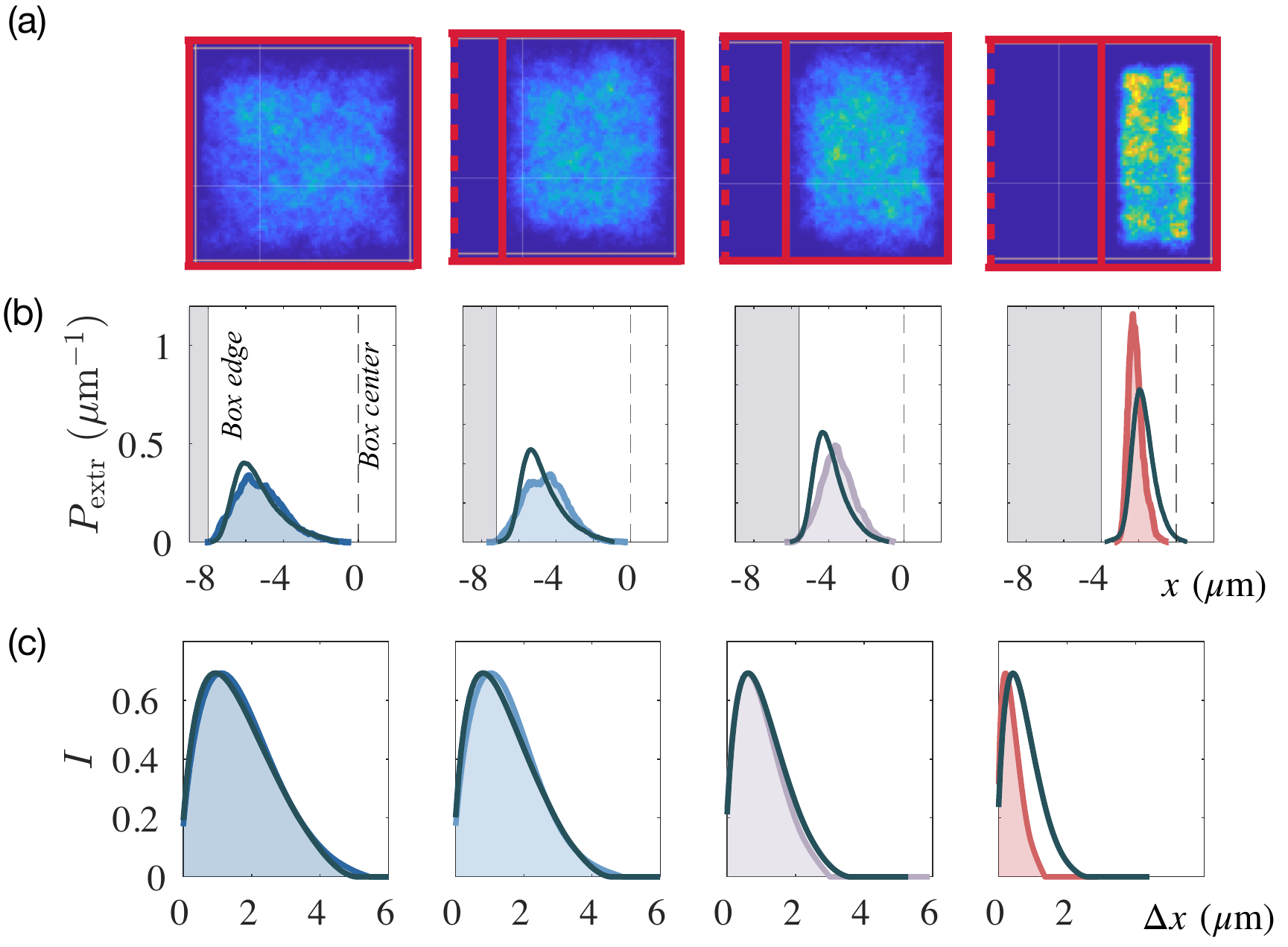}
	\caption{Main statistical properties of the engine, in 4 different compression stages (full, $20\%$, $30\%$ and $50\%$-compressed) from blue on the left to red on the right.
    (a) Measured equilibrium positional density of all particles at all times.
    (b) Probability density of the leftmost particle $P_{\rm extr}(x)$ as a function of the $x$-axis position in the box, for experimental data (blue, to red solid line and filled area) and for numerical simulations (solid gray lines).
    (c) Shannon information $I$ corresponding to a single measurement of the area $L_y \Delta x$.}
    \label{fig:Stat}
\end{figure}

The information gained by a single measurement is the Shannon entropy of this set of events $I = -p_0 \ln p_0 - p_1 \ln p_1.$
Since $p_1$ determines the probability that, upon measurement, a step of compression is applied.
It means that, on average, a number $p_1^{-1}$ of measurements are necessary to perform one compression step.
This leads to an evaluation of the mean information per compression step $\langle I \rangle = I / p_1$.
Experimentally, $p_1$ is determined, as explained in the main text, via measurements of the position of the leftmost particle.
The latter time-integrated distribution $P_{\rm extr}(x)$ is shown in Fig.~\ref{fig:Stat}~(b) for 4 sequential box sizes from $16\times 16~\rm{\mu m^2}$ (blue) to $8 \times 16 ~\rm{\mu m^2}$ (red).
The $x$ axis was arranged so that the walls are at $\pm L/2$, and $x=0$ is at the center of the box.
The result of numerical simulation is the same configuration is shown as a gray-blue solid line.
As clearly seen in the figure, $P_{\rm extr}(x)$ takes a sharper profile, with a faster decay, as the box gets compressed.
This fact is reflected in the decay of $p_1$ shown in Fig.~\ref{fig:Work}~(b) in the main text, for the same four cases, plotted against the size $\Delta x$ of the probed region near the leftmost wall.
The steeper decay denotes the decreasing probability of performing a step of large amplitude $\Delta x$ when the box is smaller.
The information $I(x)$ is derived from this measured probability $p_1$, and is plotted for the same cases in Fig.~\ref{fig:Stat}~(c).

\begin{figure}[htb!]
	\centering
	\includegraphics[width=\linewidth]{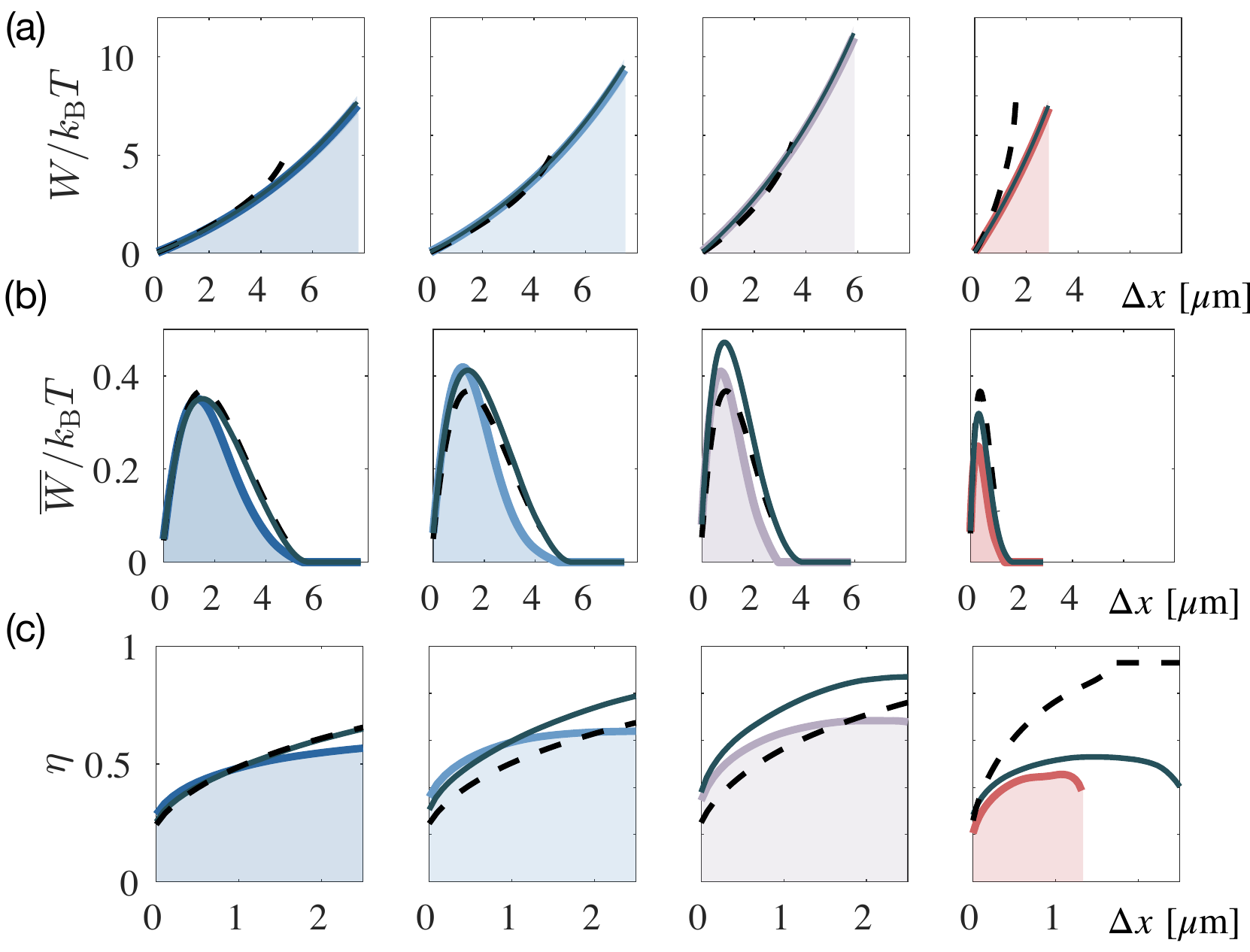}
	\caption{Energetics in 4 different compression stages (full, $20\%$, $30\%$ and $50\%$-compressed) from blue on the left to red on the right.
    (a) Work $W$ as a function of compressions $\Delta x$, for experiments (blue to red solid lines) numerical simulations (gray-blue solid lines) and analytical expression in the quasistatic regime $W = \Delta F = -k_{\rm B}T\ln(p_1)$ (black dotted line).
    (b) Mean work $\overline W= p_1 W$ in the same cases.
    (c) Information to work efficiency $\eta$ in the same cases.}
    \label{fig:Energetics}
\end{figure}

When a step of compression of amplitude $\Delta x$ is performed, the system stores an osmotic pressure which represents a work $W = \int\Pi d A$.
In the quasistatic regime, and for a steep wall, this work satisfies $W \approx \Delta F = -k_{\rm B}T\ln(p_1)$, it is shown in Fig.~\ref{fig:Energetics}~(a) for the same four cases as in Fig.~\ref{fig:Stat}.
Again, this energy gain is conditioned by $p_1$ and this allows to define a mean work per step $\overline W= p_1 W$ which is shown in Fig.~\ref{fig:Energetics}~(b) as a function of the compression $\Delta x$.
When plotted against the probability $p_1$ itself, it leads to the universal relation shown in the main text.

\begin{figure}[h!]
	\centering
	\includegraphics[width=\linewidth]{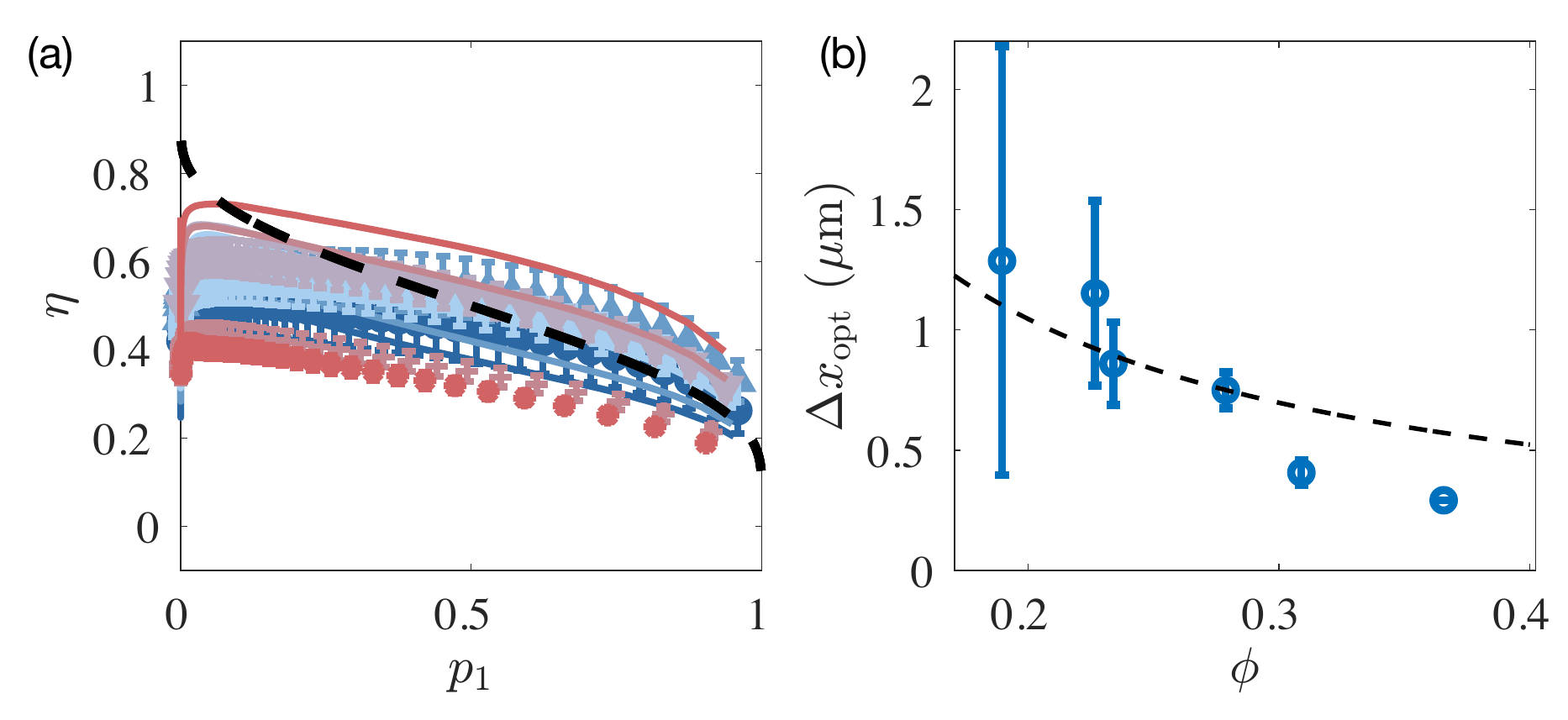}
	\caption{(a) Universal relation between the efficiency $\eta$ and the probability of a positive measurement $p_1$. The experimental results (symbols) and numerical simulations (solid lines) all fall close to the universal curve $\eta = \frac{\ln(p_1)}{(p_0/p_1) \ln(p_0)+\ln(p_1)}$ (black dash-dotted line).
    (b) Optimal step length $\Delta x_{\rm opt}$ obtained from the measured work as a function of $p_1$: the optimal step in each case corresponds to $p_1 \approx e^{-1}$ that maximizes $p_1 \ln(p_1)$.
    The dashed line corresponds to the expected $\Delta x_{\rm opt}$ for an ideal gas.}
    \label{fig:UniversalEfficiency}
\end{figure}

The information content of this single measurement, shown in Fig.~\ref{fig:Stat}~(d), combined with the work, allows to derive the information-to-work efficiency $\eta = W / k_{\rm B}T\langle I \rangle$. This quantity is shown in Fig.\ref{fig:Energetics}~(c) for the same cases as above, as a function of the compression $\Delta x$.
The universal mean work relation to $p_1$ yields an optimal operation with steps that guarantee $p_1 = e^{-1}$.
In Fig.~\ref{fig:UniversalEfficiency}~(b) we show the corresponding ideal step-size $\Delta x_{\rm opt}$ for each box size.
For an ideal gas, $p_1 = (1 - \Delta_x/L_x)^N$ with $N=8$ is the number of particle.
Hence, the optimal step-size reads $\Delta x_{\rm opt} = L_x \left( 1 - e^{-1/8}\right) = \frac{8 \pi r^2}{L_y \phi} \left( 1 - e^{-1/8}\right)$ as a function of the area fraction $\phi$.
This ideal-gas result is shown together with the measured optimal step-size in Fig.~\ref{fig:UniversalEfficiency}~(b).


\begin{thebibliography}{41}%
\makeatletter
\providecommand \@ifxundefined [1]{%
 \@ifx{#1\undefined}
}%
\providecommand \@ifnum [1]{%
 \ifnum #1\expandafter \@firstoftwo
 \else \expandafter \@secondoftwo
 \fi
}%
\providecommand \@ifx [1]{%
 \ifx #1\expandafter \@firstoftwo
 \else \expandafter \@secondoftwo
 \fi
}%
\providecommand \natexlab [1]{#1}%
\providecommand \enquote  [1]{``#1''}%
\providecommand \bibnamefont  [1]{#1}%
\providecommand \bibfnamefont [1]{#1}%
\providecommand \citenamefont [1]{#1}%
\providecommand \href@noop [0]{\@secondoftwo}%
\providecommand \href [0]{\begingroup \@sanitize@url \@href}%
\providecommand \@href[1]{\@@startlink{#1}\@@href}%
\providecommand \@@href[1]{\endgroup#1\@@endlink}%
\providecommand \@sanitize@url [0]{\catcode `\\12\catcode `\$12\catcode
  `\&12\catcode `\#12\catcode `\^12\catcode `\_12\catcode `\%12\relax}%
\providecommand \@@startlink[1]{}%
\providecommand \@@endlink[0]{}%
\providecommand \url  [0]{\begingroup\@sanitize@url \@url }%
\providecommand \@url [1]{\endgroup\@href {#1}{\urlprefix }}%
\providecommand \urlprefix  [0]{URL }%
\providecommand \Eprint [0]{\href }%
\providecommand \doibase [0]{https://doi.org/}%
\providecommand \selectlanguage [0]{\@gobble}%
\providecommand \bibinfo  [0]{\@secondoftwo}%
\providecommand \bibfield  [0]{\@secondoftwo}%
\providecommand \translation [1]{[#1]}%
\providecommand \BibitemOpen [0]{}%
\providecommand \bibitemStop [0]{}%
\providecommand \bibitemNoStop [0]{.\EOS\space}%
\providecommand \EOS [0]{\spacefactor3000\relax}%
\providecommand \BibitemShut  [1]{\csname bibitem#1\endcsname}%
\let\auto@bib@innerbib\@empty
\bibitem [{\citenamefont {Parrondo}\ \emph {et~al.}(2015)\citenamefont
  {Parrondo}, \citenamefont {Horowitz},\ and\ \citenamefont
  {Sagawa}}]{parrondo_thermodynamics_2015}%
  \BibitemOpen
  \bibfield  {author} {\bibinfo {author} {\bibfnamefont {J.~M.~R.}\
  \bibnamefont {Parrondo}}, \bibinfo {author} {\bibfnamefont {J.~M.}\
  \bibnamefont {Horowitz}},\ and\ \bibinfo {author} {\bibfnamefont
  {T.}~\bibnamefont {Sagawa}},\ }\bibfield  {title} {\bibinfo {title}
  {Thermodynamics of information},\ }\href {https://doi.org/10.1038/nphys3230}
  {\bibfield  {journal} {\bibinfo  {journal} {Nature Physics}\ }\textbf
  {\bibinfo {volume} {11}},\ \bibinfo {pages} {131} (\bibinfo {year}
  {2015})}\BibitemShut {NoStop}%
\bibitem [{\citenamefont {Goerlich}\ \emph {et~al.}(2025)\citenamefont
  {Goerlich}, \citenamefont {Hoek}, \citenamefont {Chor}, \citenamefont
  {Rahav},\ and\ \citenamefont {Roichman}}]{Goerlich2025}%
  \BibitemOpen
  \bibfield  {author} {\bibinfo {author} {\bibfnamefont {R.}~\bibnamefont
  {Goerlich}}, \bibinfo {author} {\bibfnamefont {L.}~\bibnamefont {Hoek}},
  \bibinfo {author} {\bibfnamefont {O.}~\bibnamefont {Chor}}, \bibinfo {author}
  {\bibfnamefont {S.}~\bibnamefont {Rahav}},\ and\ \bibinfo {author}
  {\bibfnamefont {Y.}~\bibnamefont {Roichman}},\ }\bibfield  {title} {\bibinfo
  {title} {Experimental realizations of information engines: beyond proof of
  concept},\ }\href {https://doi.org/10.1209/0295-5075/adbb17} {\bibfield
  {journal} {\bibinfo  {journal} {Europhysics Letters}\ }\textbf {\bibinfo
  {volume} {149}},\ \bibinfo {pages} {61001} (\bibinfo {year}
  {2025})}\BibitemShut {NoStop}%
\bibitem [{\citenamefont {Rold\'an}\ \emph {et~al.}(2014)\citenamefont
  {Rold\'an}, \citenamefont {Mart\'inez}, \citenamefont {Parrondo},\ and\
  \citenamefont {Petrov}}]{roldan_universal_2014}%
  \BibitemOpen
  \bibfield  {author} {\bibinfo {author} {\bibfnamefont {E.}~\bibnamefont
  {Rold\'an}}, \bibinfo {author} {\bibfnamefont {I.~A.}\ \bibnamefont
  {Mart\'inez}}, \bibinfo {author} {\bibfnamefont {J.~M.~R.}\ \bibnamefont
  {Parrondo}},\ and\ \bibinfo {author} {\bibfnamefont {D.}~\bibnamefont
  {Petrov}},\ }\bibfield  {title} {\bibinfo {title} {Universal features in the
  energetics of symmetry breaking},\ }\href {https://doi.org/10.1038/nphys2940}
  {\bibfield  {journal} {\bibinfo  {journal} {Nature Physics}\ }\textbf
  {\bibinfo {volume} {10}},\ \bibinfo {pages} {457} (\bibinfo {year}
  {2014})}\BibitemShut {NoStop}%
\bibitem [{\citenamefont {Koski}\ \emph {et~al.}(2014)\citenamefont {Koski},
  \citenamefont {Maisi}, \citenamefont {Sagawa},\ and\ \citenamefont
  {Pekola}}]{koski_experimental_2014}%
  \BibitemOpen
  \bibfield  {author} {\bibinfo {author} {\bibfnamefont {J.~V.}\ \bibnamefont
  {Koski}}, \bibinfo {author} {\bibfnamefont {V.~F.}\ \bibnamefont {Maisi}},
  \bibinfo {author} {\bibfnamefont {T.}~\bibnamefont {Sagawa}},\ and\ \bibinfo
  {author} {\bibfnamefont {J.~P.}\ \bibnamefont {Pekola}},\ }\bibfield  {title}
  {\bibinfo {title} {Experimental observation of the role of mutual information
  in the nonequilibrium dynamics of a {Maxwell} demon},\ }\href
  {https://doi.org/10.1103/PhysRevLett.113.030601} {\bibfield  {journal}
  {\bibinfo  {journal} {Physical Review Letters}\ }\textbf {\bibinfo {volume}
  {113}},\ \bibinfo {pages} {030601} (\bibinfo {year} {2014})}\BibitemShut
  {NoStop}%
\bibitem [{\citenamefont {Ribezzi-Crivellari}\ and\ \citenamefont
  {Ritort}(2019)}]{ribezzi-crivellari_large_2019}%
  \BibitemOpen
  \bibfield  {author} {\bibinfo {author} {\bibfnamefont {M.}~\bibnamefont
  {Ribezzi-Crivellari}}\ and\ \bibinfo {author} {\bibfnamefont
  {F.}~\bibnamefont {Ritort}},\ }\bibfield  {title} {\bibinfo {title} {Large
  work extraction and the {Landauer} limit in a continuous {Maxwell} demon},\
  }\href {https://doi.org/10.1038/s41567-019-0481-0} {\bibfield  {journal}
  {\bibinfo  {journal} {Nature Physics}\ }\textbf {\bibinfo {volume} {15}},\
  \bibinfo {pages} {660} (\bibinfo {year} {2019})}\BibitemShut {NoStop}%
\bibitem [{\citenamefont {Landauer}(1961)}]{landauer1961irreversibility}%
  \BibitemOpen
  \bibfield  {author} {\bibinfo {author} {\bibfnamefont {R.}~\bibnamefont
  {Landauer}},\ }\bibfield  {title} {\bibinfo {title} {Irreversibility and heat
  generation in the computing process},\ }\href
  {https://doi.org/10.1147/rd.53.0183} {\bibfield  {journal} {\bibinfo
  {journal} {IBM journal of research and development}\ }\textbf {\bibinfo
  {volume} {5}},\ \bibinfo {pages} {183} (\bibinfo {year} {1961})}\BibitemShut
  {NoStop}%
\bibitem [{\citenamefont {Lutz}\ and\ \citenamefont
  {Ciliberto}(2015)}]{lutz2015information}%
  \BibitemOpen
  \bibfield  {author} {\bibinfo {author} {\bibfnamefont {E.}~\bibnamefont
  {Lutz}}\ and\ \bibinfo {author} {\bibfnamefont {S.}~\bibnamefont
  {Ciliberto}},\ }\bibfield  {title} {\bibinfo {title} {Information: From
  {Maxwell}’s demon to {Landauer}’s eraser},\ }\href
  {https://doi.org/10.1063/PT.3.2912} {\bibfield  {journal} {\bibinfo
  {journal} {Physics Today}\ }\textbf {\bibinfo {volume} {68}},\ \bibinfo
  {pages} {30} (\bibinfo {year} {2015})}\BibitemShut {NoStop}%
\bibitem [{\citenamefont {Esposito}\ and\ \citenamefont {Van~den
  Broeck}(2011)}]{esposito2011second}%
  \BibitemOpen
  \bibfield  {author} {\bibinfo {author} {\bibfnamefont {M.}~\bibnamefont
  {Esposito}}\ and\ \bibinfo {author} {\bibfnamefont {C.}~\bibnamefont {Van~den
  Broeck}},\ }\bibfield  {title} {\bibinfo {title} {Second law and {Landauer}
  principle far from equilibrium},\ }\href
  {https://doi.org/10.1209/0295-5075/95/40004} {\bibfield  {journal} {\bibinfo
  {journal} {Europhysics Letters}\ }\textbf {\bibinfo {volume} {95}},\ \bibinfo
  {pages} {40004} (\bibinfo {year} {2011})}\BibitemShut {NoStop}%
\bibitem [{\citenamefont {Sagawa}\ and\ \citenamefont
  {Ueda}(2010)}]{sagawa_generalized_2010}%
  \BibitemOpen
  \bibfield  {author} {\bibinfo {author} {\bibfnamefont {T.}~\bibnamefont
  {Sagawa}}\ and\ \bibinfo {author} {\bibfnamefont {M.}~\bibnamefont {Ueda}},\
  }\bibfield  {title} {\bibinfo {title} {Generalized {Jarzynski} equality under
  nonequilibrium {Feedback} {Control}},\ }\href
  {https://doi.org/10.1103/PhysRevLett.104.090602} {\bibfield  {journal}
  {\bibinfo  {journal} {Physical Review Letters}\ }\textbf {\bibinfo {volume}
  {104}},\ \bibinfo {pages} {090602} (\bibinfo {year} {2010})}\BibitemShut
  {NoStop}%
\bibitem [{\citenamefont {Horowitz}\ and\ \citenamefont
  {Vaikuntanathan}(2010)}]{horowitz_nonequilibrium_2010}%
  \BibitemOpen
  \bibfield  {author} {\bibinfo {author} {\bibfnamefont {J.~M.}\ \bibnamefont
  {Horowitz}}\ and\ \bibinfo {author} {\bibfnamefont {S.}~\bibnamefont
  {Vaikuntanathan}},\ }\bibfield  {title} {\bibinfo {title} {Nonequilibrium
  detailed fluctuation theorem for repeated discrete feedback},\ }\href
  {https://doi.org/10.1103/PhysRevE.82.061120} {\bibfield  {journal} {\bibinfo
  {journal} {Physical Review E}\ }\textbf {\bibinfo {volume} {82}},\ \bibinfo
  {pages} {061120} (\bibinfo {year} {2010})}\BibitemShut {NoStop}%
\bibitem [{\citenamefont {Horowitz}\ and\ \citenamefont
  {Esposito}(2014)}]{horowitz2014thermodynamics}%
  \BibitemOpen
  \bibfield  {author} {\bibinfo {author} {\bibfnamefont {J.~M.}\ \bibnamefont
  {Horowitz}}\ and\ \bibinfo {author} {\bibfnamefont {M.}~\bibnamefont
  {Esposito}},\ }\bibfield  {title} {\bibinfo {title} {Thermodynamics with
  continuous information flow},\ }\href
  {https://doi.org/https://doi.org/10.1103/PhysRevX.4.031015} {\bibfield
  {journal} {\bibinfo  {journal} {Physical Review X}\ }\textbf {\bibinfo
  {volume} {4}},\ \bibinfo {pages} {031015} (\bibinfo {year}
  {2014})}\BibitemShut {NoStop}%
\bibitem [{\citenamefont {Khadka}\ \emph {et~al.}(2018)\citenamefont {Khadka},
  \citenamefont {Holubec}, \citenamefont {Yang},\ and\ \citenamefont
  {Cichos}}]{khadka2018active}%
  \BibitemOpen
  \bibfield  {author} {\bibinfo {author} {\bibfnamefont {U.}~\bibnamefont
  {Khadka}}, \bibinfo {author} {\bibfnamefont {V.}~\bibnamefont {Holubec}},
  \bibinfo {author} {\bibfnamefont {H.}~\bibnamefont {Yang}},\ and\ \bibinfo
  {author} {\bibfnamefont {F.}~\bibnamefont {Cichos}},\ }\bibfield  {title}
  {\bibinfo {title} {Active particles bound by information flows},\ }\href
  {https://doi.org/https://doi.org/10.1038/s41467-018-06445-1} {\bibfield
  {journal} {\bibinfo  {journal} {Nature communications}\ }\textbf {\bibinfo
  {volume} {9}},\ \bibinfo {pages} {3864} (\bibinfo {year} {2018})}\BibitemShut
  {NoStop}%
\bibitem [{\citenamefont {Hartich}\ \emph {et~al.}(2014)\citenamefont
  {Hartich}, \citenamefont {Barato},\ and\ \citenamefont
  {Seifert}}]{hartich2014stochastic}%
  \BibitemOpen
  \bibfield  {author} {\bibinfo {author} {\bibfnamefont {D.}~\bibnamefont
  {Hartich}}, \bibinfo {author} {\bibfnamefont {A.~C.}\ \bibnamefont
  {Barato}},\ and\ \bibinfo {author} {\bibfnamefont {U.}~\bibnamefont
  {Seifert}},\ }\bibfield  {title} {\bibinfo {title} {Stochastic thermodynamics
  of bipartite systems: transfer entropy inequalities and a {Maxwell}’s demon
  interpretation},\ }\href {https://doi.org/10.1088/1742-5468/2014/02/P02016}
  {\bibfield  {journal} {\bibinfo  {journal} {Journal of Statistical Mechanics:
  Theory and Experiment}\ }\textbf {\bibinfo {volume} {2014}},\ \bibinfo
  {pages} {P02016} (\bibinfo {year} {2014})}\BibitemShut {NoStop}%
\bibitem [{\citenamefont {Ariga}\ \emph {et~al.}(2021)\citenamefont {Ariga},
  \citenamefont {Tateishi}, \citenamefont {Tomishige},\ and\ \citenamefont
  {Mizuno}}]{ariga2021noise}%
  \BibitemOpen
  \bibfield  {author} {\bibinfo {author} {\bibfnamefont {T.}~\bibnamefont
  {Ariga}}, \bibinfo {author} {\bibfnamefont {K.}~\bibnamefont {Tateishi}},
  \bibinfo {author} {\bibfnamefont {M.}~\bibnamefont {Tomishige}},\ and\
  \bibinfo {author} {\bibfnamefont {D.}~\bibnamefont {Mizuno}},\ }\bibfield
  {title} {\bibinfo {title} {Noise-induced acceleration of single molecule
  kinesin-1},\ }\href
  {https://doi.org/https://doi.org/10.1103/PhysRevLett.127.178101} {\bibfield
  {journal} {\bibinfo  {journal} {Physical review letters}\ }\textbf {\bibinfo
  {volume} {127}},\ \bibinfo {pages} {178101} (\bibinfo {year}
  {2021})}\BibitemShut {NoStop}%
\bibitem [{\citenamefont {Buisson}\ \emph {et~al.}(2025)\citenamefont
  {Buisson}, \citenamefont {Ehrich}, \citenamefont {Leighton}, \citenamefont
  {Kundu}, \citenamefont {Saha}, \citenamefont {Bechhoefer},\ and\
  \citenamefont {Sivak}}]{buisson2025hunting}%
  \BibitemOpen
  \bibfield  {author} {\bibinfo {author} {\bibfnamefont {J.~d.}\ \bibnamefont
  {Buisson}}, \bibinfo {author} {\bibfnamefont {J.}~\bibnamefont {Ehrich}},
  \bibinfo {author} {\bibfnamefont {M.~P.}\ \bibnamefont {Leighton}}, \bibinfo
  {author} {\bibfnamefont {A.}~\bibnamefont {Kundu}}, \bibinfo {author}
  {\bibfnamefont {T.~K.}\ \bibnamefont {Saha}}, \bibinfo {author}
  {\bibfnamefont {J.}~\bibnamefont {Bechhoefer}},\ and\ \bibinfo {author}
  {\bibfnamefont {D.~A.}\ \bibnamefont {Sivak}},\ }\bibfield  {title} {\bibinfo
  {title} {Hunting for {Maxwell}'s demon in the wild},\ }\bibfield  {journal}
  {\bibinfo  {journal} {arXiv preprint}\ }\href
  {https://doi.org/10.48550/arXiv.2504.11329} {10.48550/arXiv.2504.11329}
  (\bibinfo {year} {2025})\BibitemShut {NoStop}%
\bibitem [{\citenamefont {Toyabe}\ \emph {et~al.}(2010)\citenamefont {Toyabe},
  \citenamefont {Sagawa}, \citenamefont {Ueda}, \citenamefont {Muneyuki},\ and\
  \citenamefont {Sano}}]{toyabe_experimental_2010}%
  \BibitemOpen
  \bibfield  {author} {\bibinfo {author} {\bibfnamefont {S.}~\bibnamefont
  {Toyabe}}, \bibinfo {author} {\bibfnamefont {T.}~\bibnamefont {Sagawa}},
  \bibinfo {author} {\bibfnamefont {M.}~\bibnamefont {Ueda}}, \bibinfo {author}
  {\bibfnamefont {E.}~\bibnamefont {Muneyuki}},\ and\ \bibinfo {author}
  {\bibfnamefont {M.}~\bibnamefont {Sano}},\ }\bibfield  {title} {\bibinfo
  {title} {Experimental demonstration of information-to-energy conversion and
  validation of the generalized {Jarzynski} equality},\ }\href
  {https://doi.org/10.1038/nphys1821} {\bibfield  {journal} {\bibinfo
  {journal} {Nature Physics}\ }\textbf {\bibinfo {volume} {6}},\ \bibinfo
  {pages} {988} (\bibinfo {year} {2010})}\BibitemShut {NoStop}%
\bibitem [{\citenamefont {Archambault}\ \emph {et~al.}(2024)\citenamefont
  {Archambault}, \citenamefont {Crauste-Thibierge}, \citenamefont {Ciliberto},\
  and\ \citenamefont {Bellon}}]{archambault2024inertial}%
  \BibitemOpen
  \bibfield  {author} {\bibinfo {author} {\bibfnamefont {A.}~\bibnamefont
  {Archambault}}, \bibinfo {author} {\bibfnamefont {C.}~\bibnamefont
  {Crauste-Thibierge}}, \bibinfo {author} {\bibfnamefont {S.}~\bibnamefont
  {Ciliberto}},\ and\ \bibinfo {author} {\bibfnamefont {L.}~\bibnamefont
  {Bellon}},\ }\bibfield  {title} {\bibinfo {title} {Inertial effects in
  discrete sampling information engines},\ }\href
  {https://doi.org/10.1209/0295-5075/ad8bf0} {\bibfield  {journal} {\bibinfo
  {journal} {Europhysics Letters}\ }\textbf {\bibinfo {volume} {148}},\
  \bibinfo {pages} {41002} (\bibinfo {year} {2024})}\BibitemShut {NoStop}%
\bibitem [{\citenamefont {Archambault}\ \emph {et~al.}(2025)\citenamefont
  {Archambault}, \citenamefont {Crauste-Thibierge}, \citenamefont {Imparato},
  \citenamefont {Jarzynski}, \citenamefont {Ciliberto},\ and\ \citenamefont
  {Bellon}}]{archambault2025}%
  \BibitemOpen
  \bibfield  {author} {\bibinfo {author} {\bibfnamefont {A.}~\bibnamefont
  {Archambault}}, \bibinfo {author} {\bibfnamefont {C.}~\bibnamefont
  {Crauste-Thibierge}}, \bibinfo {author} {\bibfnamefont {A.}~\bibnamefont
  {Imparato}}, \bibinfo {author} {\bibfnamefont {C.}~\bibnamefont {Jarzynski}},
  \bibinfo {author} {\bibfnamefont {S.}~\bibnamefont {Ciliberto}},\ and\
  \bibinfo {author} {\bibfnamefont {L.}~\bibnamefont {Bellon}},\ }\bibfield
  {title} {\bibinfo {title} {Information engine fueled by first-passage
  times},\ }\href {https://doi.org/https://doi.org/10.1103/s9kj-lczm}
  {\bibfield  {journal} {\bibinfo  {journal} {Phys. Rev. Lett.}\ }\textbf
  {\bibinfo {volume} {135}},\ \bibinfo {pages} {147101} (\bibinfo {year}
  {2025})}\BibitemShut {NoStop}%
\bibitem [{\citenamefont {Lagoin}\ \emph {et~al.}(2022)\citenamefont {Lagoin},
  \citenamefont {Crauste-Thibierge},\ and\ \citenamefont
  {Naert}}]{lagoin2022human}%
  \BibitemOpen
  \bibfield  {author} {\bibinfo {author} {\bibfnamefont {M.}~\bibnamefont
  {Lagoin}}, \bibinfo {author} {\bibfnamefont {C.}~\bibnamefont
  {Crauste-Thibierge}},\ and\ \bibinfo {author} {\bibfnamefont
  {A.}~\bibnamefont {Naert}},\ }\bibfield  {title} {\bibinfo {title}
  {Human-scale {Brownian} ratchet: a historical thought experiment},\ }\href
  {https://doi.org/https://doi.org/10.1103/PhysRevLett.129.120606} {\bibfield
  {journal} {\bibinfo  {journal} {Physical Review Letters}\ }\textbf {\bibinfo
  {volume} {129}},\ \bibinfo {pages} {120606} (\bibinfo {year}
  {2022})}\BibitemShut {NoStop}%
\bibitem [{\citenamefont {Chor}\ \emph {et~al.}(2023)\citenamefont {Chor},
  \citenamefont {Sohachi}, \citenamefont {Goerlich}, \citenamefont {Rosen},
  \citenamefont {Rahav},\ and\ \citenamefont {Roichman}}]{chor2023many}%
  \BibitemOpen
  \bibfield  {author} {\bibinfo {author} {\bibfnamefont {O.}~\bibnamefont
  {Chor}}, \bibinfo {author} {\bibfnamefont {A.}~\bibnamefont {Sohachi}},
  \bibinfo {author} {\bibfnamefont {R.}~\bibnamefont {Goerlich}}, \bibinfo
  {author} {\bibfnamefont {E.}~\bibnamefont {Rosen}}, \bibinfo {author}
  {\bibfnamefont {S.}~\bibnamefont {Rahav}},\ and\ \bibinfo {author}
  {\bibfnamefont {Y.}~\bibnamefont {Roichman}},\ }\bibfield  {title} {\bibinfo
  {title} {Many-body {Szilard} engine with giant number fluctuations},\ }\href
  {https://doi.org/https://doi.org/10.1103/PhysRevResearch.5.043193} {\bibfield
   {journal} {\bibinfo  {journal} {Physical Review Research}\ }\textbf
  {\bibinfo {volume} {5}},\ \bibinfo {pages} {043193} (\bibinfo {year}
  {2023})}\BibitemShut {NoStop}%
\bibitem [{\citenamefont {Howard}\ and\ \citenamefont
  {Clark}(2002)}]{howard2002mechanics}%
  \BibitemOpen
  \bibfield  {author} {\bibinfo {author} {\bibfnamefont {J.}~\bibnamefont
  {Howard}}\ and\ \bibinfo {author} {\bibfnamefont {R.}~\bibnamefont {Clark}},\
  }\bibfield  {title} {\bibinfo {title} {Mechanics of motor proteins and the
  cytoskeleton},\ }\href@noop {} {\bibfield  {journal} {\bibinfo  {journal}
  {Appl. Mech. Rev.}\ }\textbf {\bibinfo {volume} {55}},\ \bibinfo {pages}
  {B39} (\bibinfo {year} {2002})}\BibitemShut {NoStop}%
\bibitem [{\citenamefont {Maxwell}(1871)}]{maxwell1871theory}%
  \BibitemOpen
  \bibfield  {author} {\bibinfo {author} {\bibfnamefont {J.~C.}\ \bibnamefont
  {Maxwell}},\ }\bibfield  {title} {\bibinfo {title} {Theory of heat},\
  }\href@noop {} {\bibfield  {journal} {\bibinfo  {journal} {London, UK:
  Longmans}\ } (\bibinfo {year} {1871})}\BibitemShut {NoStop}%
\bibitem [{\citenamefont {Helfand}\ \emph {et~al.}(1961)\citenamefont
  {Helfand}, \citenamefont {Frisch},\ and\ \citenamefont
  {Lebowitz}}]{helfand1961theory}%
  \BibitemOpen
  \bibfield  {author} {\bibinfo {author} {\bibfnamefont {E.}~\bibnamefont
  {Helfand}}, \bibinfo {author} {\bibfnamefont {H.}~\bibnamefont {Frisch}},\
  and\ \bibinfo {author} {\bibfnamefont {J.}~\bibnamefont {Lebowitz}},\
  }\bibfield  {title} {\bibinfo {title} {Theory of the two-and one-dimensional
  rigid sphere fluids},\ }\href
  {https://doi.org/https://doi.org/10.1063/1.1731629} {\bibfield  {journal}
  {\bibinfo  {journal} {The Journal of Chemical Physics}\ }\textbf {\bibinfo
  {volume} {34}},\ \bibinfo {pages} {1037} (\bibinfo {year}
  {1961})}\BibitemShut {NoStop}%
\bibitem [{\citenamefont {Thorneywork}\ \emph {et~al.}(2017)\citenamefont
  {Thorneywork}, \citenamefont {Abbott}, \citenamefont {Aarts},\ and\
  \citenamefont {Dullens}}]{thorneywork2017two}%
  \BibitemOpen
  \bibfield  {author} {\bibinfo {author} {\bibfnamefont {A.~L.}\ \bibnamefont
  {Thorneywork}}, \bibinfo {author} {\bibfnamefont {J.~L.}\ \bibnamefont
  {Abbott}}, \bibinfo {author} {\bibfnamefont {D.~G.}\ \bibnamefont {Aarts}},\
  and\ \bibinfo {author} {\bibfnamefont {R.~P.}\ \bibnamefont {Dullens}},\
  }\bibfield  {title} {\bibinfo {title} {Two-dimensional melting of colloidal
  hard spheres},\ }\href
  {https://doi.org/https://doi.org/10.1103/PhysRevLett.118.158001} {\bibfield
  {journal} {\bibinfo  {journal} {Physical review letters}\ }\textbf {\bibinfo
  {volume} {118}},\ \bibinfo {pages} {158001} (\bibinfo {year}
  {2017})}\BibitemShut {NoStop}%
\bibitem [{\citenamefont {Royall}\ \emph {et~al.}(2024)\citenamefont {Royall},
  \citenamefont {Charbonneau}, \citenamefont {Dijkstra}, \citenamefont {Russo},
  \citenamefont {Smallenburg}, \citenamefont {Speck},\ and\ \citenamefont
  {Valeriani}}]{royall2024colloidal}%
  \BibitemOpen
  \bibfield  {author} {\bibinfo {author} {\bibfnamefont {C.~P.}\ \bibnamefont
  {Royall}}, \bibinfo {author} {\bibfnamefont {P.}~\bibnamefont {Charbonneau}},
  \bibinfo {author} {\bibfnamefont {M.}~\bibnamefont {Dijkstra}}, \bibinfo
  {author} {\bibfnamefont {J.}~\bibnamefont {Russo}}, \bibinfo {author}
  {\bibfnamefont {F.}~\bibnamefont {Smallenburg}}, \bibinfo {author}
  {\bibfnamefont {T.}~\bibnamefont {Speck}},\ and\ \bibinfo {author}
  {\bibfnamefont {C.}~\bibnamefont {Valeriani}},\ }\bibfield  {title} {\bibinfo
  {title} {Colloidal hard spheres: Triumphs, challenges, and mysteries},\
  }\href {https://doi.org/https://doi.org/10.1103/RevModPhys.96.045003}
  {\bibfield  {journal} {\bibinfo  {journal} {Reviews of Modern Physics}\
  }\textbf {\bibinfo {volume} {96}},\ \bibinfo {pages} {045003} (\bibinfo
  {year} {2024})}\BibitemShut {NoStop}%
\bibitem [{\citenamefont {Shannon}(1948)}]{Shannon1949}%
  \BibitemOpen
  \bibfield  {author} {\bibinfo {author} {\bibfnamefont {C.~E.}\ \bibnamefont
  {Shannon}},\ }\bibfield  {title} {\bibinfo {title} {A mathematical theory of
  communication},\ }\href {https://doi.org/10.1002/j.1538-7305.1948.tb01338.x}
  {\bibfield  {journal} {\bibinfo  {journal} {The Bell system technical
  journal}\ }\textbf {\bibinfo {volume} {27}},\ \bibinfo {pages} {379}
  (\bibinfo {year} {1948})}\BibitemShut {NoStop}%
\bibitem [{\citenamefont {Maruyama}\ \emph {et~al.}(2009)\citenamefont
  {Maruyama}, \citenamefont {Nori},\ and\ \citenamefont
  {Vedral}}]{maruyama_colloquium_2009}%
  \BibitemOpen
  \bibfield  {author} {\bibinfo {author} {\bibfnamefont {K.}~\bibnamefont
  {Maruyama}}, \bibinfo {author} {\bibfnamefont {F.}~\bibnamefont {Nori}},\
  and\ \bibinfo {author} {\bibfnamefont {V.}~\bibnamefont {Vedral}},\
  }\bibfield  {title} {\bibinfo {title} {\textit{{Colloquium}} : {The} physics
  of {Maxwell}’s demon and information},\ }\href
  {https://doi.org/10.1103/RevModPhys.81.1} {\bibfield  {journal} {\bibinfo
  {journal} {Reviews of Modern Physics}\ }\textbf {\bibinfo {volume} {81}},\
  \bibinfo {pages} {1} (\bibinfo {year} {2009})}\BibitemShut {NoStop}%
\bibitem [{\citenamefont {Jarzynski}(2012)}]{jarzynski2012equalities}%
  \BibitemOpen
  \bibfield  {author} {\bibinfo {author} {\bibfnamefont {C.}~\bibnamefont
  {Jarzynski}},\ }\bibfield  {title} {\bibinfo {title} {Equalities and
  inequalities: Irreversibility and the second law of thermodynamics at the
  nanoscale},\ }in\ \href
  {https://doi.org/10.1146/annurev-conmatphys-062910-140506} {\emph {\bibinfo
  {booktitle} {Time: Poincar{\'e} Seminar 2010}}}\ (\bibinfo {organization}
  {Springer},\ \bibinfo {year} {2012})\ pp.\ \bibinfo {pages}
  {145--172}\BibitemShut {NoStop}%
\bibitem [{\citenamefont {Gavrilov}\ \emph {et~al.}(2017)\citenamefont
  {Gavrilov}, \citenamefont {Ch{\'e}trite},\ and\ \citenamefont
  {Bechhoefer}}]{gavrilov2017direct}%
  \BibitemOpen
  \bibfield  {author} {\bibinfo {author} {\bibfnamefont {M.}~\bibnamefont
  {Gavrilov}}, \bibinfo {author} {\bibfnamefont {R.}~\bibnamefont
  {Ch{\'e}trite}},\ and\ \bibinfo {author} {\bibfnamefont {J.}~\bibnamefont
  {Bechhoefer}},\ }\bibfield  {title} {\bibinfo {title} {Direct measurement of
  weakly nonequilibrium system entropy is consistent with {Gibbs}-{Shannon}
  form},\ }\href {https://doi.org/https://doi.org/10.1073/pnas.1708689114}
  {\bibfield  {journal} {\bibinfo  {journal} {Proceedings of the National
  Academy of Sciences}\ }\textbf {\bibinfo {volume} {114}},\ \bibinfo {pages}
  {11097} (\bibinfo {year} {2017})}\BibitemShut {NoStop}%
\bibitem [{\citenamefont {Sekimoto}(1998)}]{sekimoto1998langevin}%
  \BibitemOpen
  \bibfield  {author} {\bibinfo {author} {\bibfnamefont {K.}~\bibnamefont
  {Sekimoto}},\ }\bibfield  {title} {\bibinfo {title} {Langevin equation and
  thermodynamics},\ }\href
  {https://doi.org/https://doi.org/10.1143/PTPS.130.17} {\bibfield  {journal}
  {\bibinfo  {journal} {Progress of Theoretical Physics Supplement}\ }\textbf
  {\bibinfo {volume} {130}},\ \bibinfo {pages} {17} (\bibinfo {year}
  {1998})}\BibitemShut {NoStop}%
\bibitem [{\citenamefont {Saha}\ \emph {et~al.}(2021)\citenamefont {Saha},
  \citenamefont {Lucero}, \citenamefont {Ehrich}, \citenamefont {Sivak},\ and\
  \citenamefont {Bechhoefer}}]{saha_maximizing_2021}%
  \BibitemOpen
  \bibfield  {author} {\bibinfo {author} {\bibfnamefont {T.~K.}\ \bibnamefont
  {Saha}}, \bibinfo {author} {\bibfnamefont {J.~N.~E.}\ \bibnamefont {Lucero}},
  \bibinfo {author} {\bibfnamefont {J.}~\bibnamefont {Ehrich}}, \bibinfo
  {author} {\bibfnamefont {D.~A.}\ \bibnamefont {Sivak}},\ and\ \bibinfo
  {author} {\bibfnamefont {J.}~\bibnamefont {Bechhoefer}},\ }\bibfield  {title}
  {\bibinfo {title} {Maximizing power and velocity of an information engine},\
  }\href {https://doi.org/10.1073/pnas.2023356118} {\bibfield  {journal}
  {\bibinfo  {journal} {Proceedings of the National Academy of Sciences}\
  }\textbf {\bibinfo {volume} {118}},\ \bibinfo {pages} {e2023356118} (\bibinfo
  {year} {2021})}\BibitemShut {NoStop}%
\bibitem [{\citenamefont {Admon}\ \emph {et~al.}(2018)\citenamefont {Admon},
  \citenamefont {Rahav},\ and\ \citenamefont
  {Roichman}}]{admon_experimental_2018}%
  \BibitemOpen
  \bibfield  {author} {\bibinfo {author} {\bibfnamefont {T.}~\bibnamefont
  {Admon}}, \bibinfo {author} {\bibfnamefont {S.}~\bibnamefont {Rahav}},\ and\
  \bibinfo {author} {\bibfnamefont {Y.}~\bibnamefont {Roichman}},\ }\bibfield
  {title} {\bibinfo {title} {Experimental realization of an information machine
  with tunable temporal correlations},\ }\href
  {https://doi.org/10.1103/PhysRevLett.121.180601} {\bibfield  {journal}
  {\bibinfo  {journal} {Physical Review Letters}\ }\textbf {\bibinfo {volume}
  {121}},\ \bibinfo {pages} {180601} (\bibinfo {year} {2018})}\BibitemShut
  {NoStop}%
\bibitem [{\citenamefont {Szilard}(1929)}]{szilard1929}%
  \BibitemOpen
  \bibfield  {author} {\bibinfo {author} {\bibfnamefont {L.}~\bibnamefont
  {Szilard}},\ }\bibfield  {title} {\bibinfo {title} {{\"U}ber die
  {Entropieverminderung} in einem thermodynamischen {System} bei {Eingriffen}
  intelligenter {Wesen}},\ }\href
  {https://doi.org/https://doi.org/10.1007/BF01341281} {\bibfield  {journal}
  {\bibinfo  {journal} {Zeitschrift f{\"u}r Physik}\ }\textbf {\bibinfo
  {volume} {53}},\ \bibinfo {pages} {840} (\bibinfo {year} {1929})}\BibitemShut
  {NoStop}%
\bibitem [{\citenamefont {Ariga}\ \emph {et~al.}(2018)\citenamefont {Ariga},
  \citenamefont {Tomishige},\ and\ \citenamefont
  {Mizuno}}]{ariga2018nonequilibrium}%
  \BibitemOpen
  \bibfield  {author} {\bibinfo {author} {\bibfnamefont {T.}~\bibnamefont
  {Ariga}}, \bibinfo {author} {\bibfnamefont {M.}~\bibnamefont {Tomishige}},\
  and\ \bibinfo {author} {\bibfnamefont {D.}~\bibnamefont {Mizuno}},\
  }\bibfield  {title} {\bibinfo {title} {Nonequilibrium energetics of molecular
  motor kinesin},\ }\href
  {https://doi.org/https://doi.org/10.1103/PhysRevLett.121.218101} {\bibfield
  {journal} {\bibinfo  {journal} {Physical review letters}\ }\textbf {\bibinfo
  {volume} {121}},\ \bibinfo {pages} {218101} (\bibinfo {year}
  {2018})}\BibitemShut {NoStop}%
\bibitem [{\citenamefont {Guo}\ \emph {et~al.}(2014)\citenamefont {Guo},
  \citenamefont {Ehrlicher}, \citenamefont {Jensen}, \citenamefont {Renz},
  \citenamefont {Moore}, \citenamefont {Goldman}, \citenamefont
  {Lippincott-Schwartz}, \citenamefont {Mackintosh},\ and\ \citenamefont
  {Weitz}}]{guo2014probing}%
  \BibitemOpen
  \bibfield  {author} {\bibinfo {author} {\bibfnamefont {M.}~\bibnamefont
  {Guo}}, \bibinfo {author} {\bibfnamefont {A.~J.}\ \bibnamefont {Ehrlicher}},
  \bibinfo {author} {\bibfnamefont {M.~H.}\ \bibnamefont {Jensen}}, \bibinfo
  {author} {\bibfnamefont {M.}~\bibnamefont {Renz}}, \bibinfo {author}
  {\bibfnamefont {J.~R.}\ \bibnamefont {Moore}}, \bibinfo {author}
  {\bibfnamefont {R.~D.}\ \bibnamefont {Goldman}}, \bibinfo {author}
  {\bibfnamefont {J.}~\bibnamefont {Lippincott-Schwartz}}, \bibinfo {author}
  {\bibfnamefont {F.~C.}\ \bibnamefont {Mackintosh}},\ and\ \bibinfo {author}
  {\bibfnamefont {D.~A.}\ \bibnamefont {Weitz}},\ }\bibfield  {title} {\bibinfo
  {title} {Probing the stochastic, motor-driven properties of the cytoplasm
  using force spectrum microscopy},\ }\href
  {https://doi.org/10.1016/j.cell.2014.06.051} {\bibfield  {journal} {\bibinfo
  {journal} {Cell}\ }\textbf {\bibinfo {volume} {158}},\ \bibinfo {pages} {822}
  (\bibinfo {year} {2014})}\BibitemShut {NoStop}%
\bibitem [{\citenamefont {Krishnamurthy}\ \emph {et~al.}(2016)\citenamefont
  {Krishnamurthy}, \citenamefont {Ghosh}, \citenamefont {Chatterji},
  \citenamefont {Ganapathy},\ and\ \citenamefont
  {Sood}}]{krishnamurthy2016micrometre}%
  \BibitemOpen
  \bibfield  {author} {\bibinfo {author} {\bibfnamefont {S.}~\bibnamefont
  {Krishnamurthy}}, \bibinfo {author} {\bibfnamefont {S.}~\bibnamefont
  {Ghosh}}, \bibinfo {author} {\bibfnamefont {D.}~\bibnamefont {Chatterji}},
  \bibinfo {author} {\bibfnamefont {R.}~\bibnamefont {Ganapathy}},\ and\
  \bibinfo {author} {\bibfnamefont {A.}~\bibnamefont {Sood}},\ }\bibfield
  {title} {\bibinfo {title} {A micrometre-sized heat engine operating between
  bacterial reservoirs},\ }\href
  {https://doi.org/https://doi.org/10.1038/nphys3870} {\bibfield  {journal}
  {\bibinfo  {journal} {Nature Physics}\ }\textbf {\bibinfo {volume} {12}},\
  \bibinfo {pages} {1134} (\bibinfo {year} {2016})}\BibitemShut {NoStop}%
\bibitem [{\citenamefont {Saha}\ \emph {et~al.}(2023)\citenamefont {Saha},
  \citenamefont {Ehrich}, \citenamefont {Gavrilov}, \citenamefont {Still},
  \citenamefont {Sivak},\ and\ \citenamefont
  {Bechhoefer}}]{saha2023information}%
  \BibitemOpen
  \bibfield  {author} {\bibinfo {author} {\bibfnamefont {T.~K.}\ \bibnamefont
  {Saha}}, \bibinfo {author} {\bibfnamefont {J.}~\bibnamefont {Ehrich}},
  \bibinfo {author} {\bibfnamefont {M.}~\bibnamefont {Gavrilov}}, \bibinfo
  {author} {\bibfnamefont {S.}~\bibnamefont {Still}}, \bibinfo {author}
  {\bibfnamefont {D.~A.}\ \bibnamefont {Sivak}},\ and\ \bibinfo {author}
  {\bibfnamefont {J.}~\bibnamefont {Bechhoefer}},\ }\bibfield  {title}
  {\bibinfo {title} {Information engine in a nonequilibrium bath},\ }\href
  {https://doi.org/https://doi.org/10.1103/PhysRevLett.131.057101} {\bibfield
  {journal} {\bibinfo  {journal} {Physical Review Letters}\ }\textbf {\bibinfo
  {volume} {131}},\ \bibinfo {pages} {057101} (\bibinfo {year}
  {2023})}\BibitemShut {NoStop}%
\bibitem [{\citenamefont {Ramaswamy}\ \emph {et~al.}(2003)\citenamefont
  {Ramaswamy}, \citenamefont {Simha},\ and\ \citenamefont
  {Toner}}]{ramaswamy2003active}%
  \BibitemOpen
  \bibfield  {author} {\bibinfo {author} {\bibfnamefont {S.}~\bibnamefont
  {Ramaswamy}}, \bibinfo {author} {\bibfnamefont {R.~A.}\ \bibnamefont
  {Simha}},\ and\ \bibinfo {author} {\bibfnamefont {J.}~\bibnamefont {Toner}},\
  }\bibfield  {title} {\bibinfo {title} {Active nematics on a substrate: Giant
  number fluctuations and long-time tails},\ }\href
  {https://doi.org/10.1209/epl/i2003-00346-7} {\bibfield  {journal} {\bibinfo
  {journal} {Europhysics Letters}\ }\textbf {\bibinfo {volume} {62}},\ \bibinfo
  {pages} {196} (\bibinfo {year} {2003})}\BibitemShut {NoStop}%
\bibitem [{\citenamefont {Crocker}\ and\ \citenamefont
  {Grier}(1996)}]{crocker1996methods}%
  \BibitemOpen
  \bibfield  {author} {\bibinfo {author} {\bibfnamefont {J.~C.}\ \bibnamefont
  {Crocker}}\ and\ \bibinfo {author} {\bibfnamefont {D.~G.}\ \bibnamefont
  {Grier}},\ }\bibfield  {title} {\bibinfo {title} {Methods of digital video
  microscopy for colloidal studies},\ }\href@noop {} {\bibfield  {journal}
  {\bibinfo  {journal} {Journal of colloid and interface science}\ }\textbf
  {\bibinfo {volume} {179}},\ \bibinfo {pages} {298} (\bibinfo {year}
  {1996})}\BibitemShut {NoStop}%
\bibitem [{\citenamefont {Volpe}\ and\ \citenamefont
  {Volpe}(2013)}]{volpe_simulation_2013}%
  \BibitemOpen
  \bibfield  {author} {\bibinfo {author} {\bibfnamefont {G.}~\bibnamefont
  {Volpe}}\ and\ \bibinfo {author} {\bibfnamefont {G.}~\bibnamefont {Volpe}},\
  }\bibfield  {title} {\bibinfo {title} {Simulation of a {Brownian} particle in
  an optical trap},\ }\href {https://doi.org/10.1119/1.4772632} {\bibfield
  {journal} {\bibinfo  {journal} {American Journal of Physics}\ }\textbf
  {\bibinfo {volume} {81}},\ \bibinfo {pages} {224} (\bibinfo {year}
  {2013})}\BibitemShut {NoStop}%
\bibitem [{\citenamefont {Hess}\ \emph {et~al.}(1998)\citenamefont {Hess},
  \citenamefont {Kröger},\ and\ \citenamefont
  {Voigt}}]{hess_thermomechanical_1998}%
  \BibitemOpen
  \bibfield  {author} {\bibinfo {author} {\bibfnamefont {S.}~\bibnamefont
  {Hess}}, \bibinfo {author} {\bibfnamefont {M.}~\bibnamefont {Kröger}},\ and\
  \bibinfo {author} {\bibfnamefont {H.}~\bibnamefont {Voigt}},\ }\bibfield
  {title} {\bibinfo {title} {Thermomechanical properties of the
  {WCA}–{Lennard}-{Jones} model system in its fluid and solid states},\
  }\href {https://doi.org/10.1016/S0378-4371(97)00612-2} {\bibfield  {journal}
  {\bibinfo  {journal} {Physica A: Statistical Mechanics and its Applications}\
  }\textbf {\bibinfo {volume} {250}},\ \bibinfo {pages} {58} (\bibinfo {year}
  {1998})}\BibitemShut {NoStop}%
\end{thebibliography}
\end{document}